\documentclass[aps,preprint,amsmath,amssymb,t1,11pt]{revtex4}
\usepackage{graphicx}
\usepackage{epstopdf}
\usepackage[T1]{fontenc}
\usepackage{slashed}
\begin{document}
\title{Probing anomalous $tq\gamma$ and $tqg$ couplings via single top production in association with photon at FCC-hh}
\author{K.Y. Oyulmaz}
\email[]{kaan.oyulmaz@gmail.com}
\author{A. Senol}
\email[]{senol_a@ibu.edu.tr} 
\author{H. Denizli}
\email[]{denizli_h@ibu.edu.tr}
\affiliation{Department of Physics, Bolu Abant Izzet Baysal University, 14280, Bolu, Turkey}
\author{A. Yilmaz}
\email[]{aliyilmaz@giresun.edu.tr}
\affiliation{Department of Electrical and Electronics Engineering, Giresun University, 28200 Giresun, Turkey}
\author{I. Turk Cakir}
\email[]{ilkay.turk.cakir@cern.ch}
\affiliation{Department of Energy Systems Engineering, Giresun University, 28200 Giresun, Turkey}
\author{O. Cakir}
\email[]{ocakir@science.ankara.edu.tr}
\affiliation{Department of Physics, Ankara University,  06100, Ankara, Turkey}
\begin{abstract}
 We study the anomalous FCNC $tq\gamma$ and $tqg$ couplings via $pp\to Wb\gamma+X$ signal process including realistic detector effects for both leptonic and hadronic decay channels of the W boson at 100 TeV FCC-hh. The relevant backgrounds are considered in the cut based analysis to obtain not only limits on the anomalous $\lambda$ and $\zeta$ couplings but also branching ratios of $t\to q\gamma$ and $t\to qg$ decay channels. We find that the sensitivity to the branching ratio of $t \to q \gamma$ channel is three order better than the available LHC experimental limits, and it is comparable  for the branching ratio of the $t\to qg$ decay channel with an integrated luminosity of 10 ab$^{-1}$ at 2$\sigma$  significance level.
\end{abstract}

\maketitle

\section{Introduction}
One of the most sensitive probe to search for a new physics beyond the Standard Model (SM) is the top quark with mass of 173.0$\pm$ 0.4 GeV \cite{Tanabashi:2018oca} close to electroweak symmetry breaking scale. Flavor changing neutral current interactions involving a top quark, other quark flavors and neutral gauge boson are forbidden at the tree level and are suppressed in a loop level due to Glashow-Iliopoulos-Maiani mechanism \cite{Glashow70}. The predicted SM branching ratios of the top quark FCNC decays to a gluon, photon, $Z$ or Higgs boson and up-type quarks are expected to be $\mathcal{O} (10^{-12}-10^{-17})$ and are out of range for current experimental sensitivity \cite{AguilarSaavedra:2004wm}. These branching ratios significantly improved in the certain parameter space of many different models beyond the SM and are close to the current experimental limits ($\mathcal{O} (10^{-4}-10^{-5}$)). Therefore, the possible deviation from SM predictions of FCNC $tq\gamma$ and $tqg$ couplings would imply the existence of new physics beyond the SM. Recently, the exclusion limits on the top quark FCNC couplings have significantly improved by the LHC. The current experimental constraints on the branching ratio of the top quark FCNC decays obtained at the ATLAS and CMS with 95\% confidence level (C.L.) are tabulated in Table \ref{BR}.

Probe of the new physics effects on FCNC top interactions in a model independent way is the effective Lagrangian approach \cite{AguilarSaavedra:2008zc,AguilarSaavedra:2009mx}. In this approach, anomalous FCNC couplings are described by higher-dimensional effective operators independently from the underlying theory. Anomalous FCNC couplings have been extensively studied using this approach in the literature \cite{Han:1998tp,delAguila:1999kfp,Belyaev:2001hf,Alan:2002wv,Cakir:2003cg,Yang:2004af,Cakir:2005rf,Zhang:2008yn,Cakir:2009rq,Drobnak:2010by,Gao:2011fx,Billur:2013ela,Agram:2013koa,Inan:2014mua,Koksal:2014hba,Hesari:2014eua,Khanpour:2014xla,Sun:2014qoa,Goldouzian:2014nha,Degrande:2014tta,Khatibi:2014via,Khatibi:2015aal,Hesari:2015oya,Sun:2016kek,Guo:2016kea,Liu:2016dag,Goldouzian:2016mrt,Zarnecki:2017cmf,Wang:2017pdg,Denizli:2017cfx,TurkCakir:2017rvu, Cakir:2018ruj,Shen:2018mlj,Liu:2018bxa,Banerjee:2018fsx,Chala:2018agk}.

 The effective Lagrangian for the FCNC $tq\gamma$ and $tqg$ couplings can be written \cite{AguilarSaavedra:2008zc,AguilarSaavedra:2009mx}
\begin{eqnarray}\label{eq1}
 L_{FCNC}&=& \frac{g_{s}}{m_{t}}\sum_{u,c} \bar q \lambda^a\sigma^{\mu\nu}(\zeta_{qt}^LP_L+\zeta_{qt}^RP_R)tG_{\mu\nu}^a
 +\frac{g_{e}}{2m_{t}}\sum_{u,c}\bar{q}\sigma^{\mu\nu}(\lambda_{qt}^{L}P_{L}+\lambda_{qt}^{R}P_{R})tA_{\mu\nu} + h.c.
\end{eqnarray}
where $g_s$ and $g_{e}$ are the strong and the electromagnetic coupling constants, respectively; $\lambda^a$ are the Gell-Mann matrices with $a=1,...,8$. $\zeta_{qt}^{L(R)}$ and $\lambda_{qt}^{L(R)}$ are the strength of anomalous FCNC couplings for $tqg$ and $tq\gamma$ vertices, respectively; $P_{L(R)}$ denotes the left (right) handed projection operators; $\sigma^{\mu\nu}$  is the tensor defined as $\sigma^{\mu\nu}=\frac{i}{2}[\gamma^{\mu},\gamma^{\nu}]$ for the FCNC interactions. We assumed no specific chirality for the FCNC interaction vertices, i.e. $\lambda_{qt}^{L}=\lambda_{qt}^{R}=\lambda_q$ and  $\zeta_{qt}^{L}=\zeta_{qt}^{R}=\zeta_q$ in this study. 

The FCNC effects involving a top quark are phenomenologically studied in many final states with various sensitivities. Mostly anomalous FCNC couplings are investigated through FCNC decay of top quarks in the processes where large number of top quarks are produced at high energy hadron colliders.  However, this situation creates disadvantages, such as separating from generic multijet production by Quantum ChromoDynamics (QCD), especially when determining $tqg$ couplings. Direct single top production in association with a photon is suggested to be powerful probe to search for existence not only $tqg$ vertices but also  $tq\gamma$ vertices in hadron colliders. One can expect even further improvements on these bounds with a higher center of mass energy colliders. The Future Circular Collider (FCC) which has the potential to search for a wide parameter range of new physics is the energy frontier collider project currently under consideration \cite{FCC}. FCC-hh, is a unique option of FCC, has a design providing proton-proton collisions at the proposed 100 TeV centre-of-mass energy with peak luminosity $5\times10^{34}$ $cm^{-2}s^{-1}$ \cite{Mangano:2017tke}. 

In this study, we focus on both hadronic and leptonic decays of the final state W in the $pp\to W b\gamma$ signal process to investigate the anomalous FCNC $tqg$ ( $\zeta_q$) and $tq\gamma$ ( $\lambda_q$) couplings at FCC-hh. Details of event selection and cuts on kinematic variables are discussed for the signal and  relevant SM background processes in addition to SM background of the same final state with the signal process. Finally, We conclude with the prediction on the sensitivity of FCC-hh to anomalous FCNC $tqg$ ( $\zeta_q$) and $tq\gamma$ ( $\lambda_q$) couplings.

\section{Signal Cross Sections}

\begin{figure}
\includegraphics[scale=0.7]{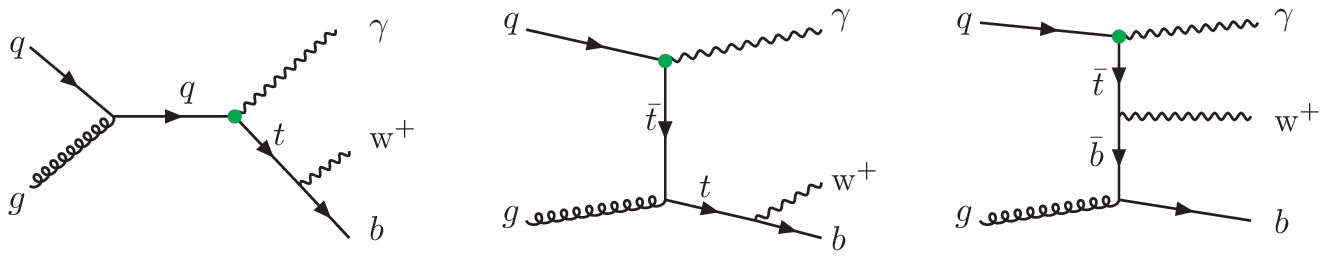}\\
\includegraphics[scale=0.8]{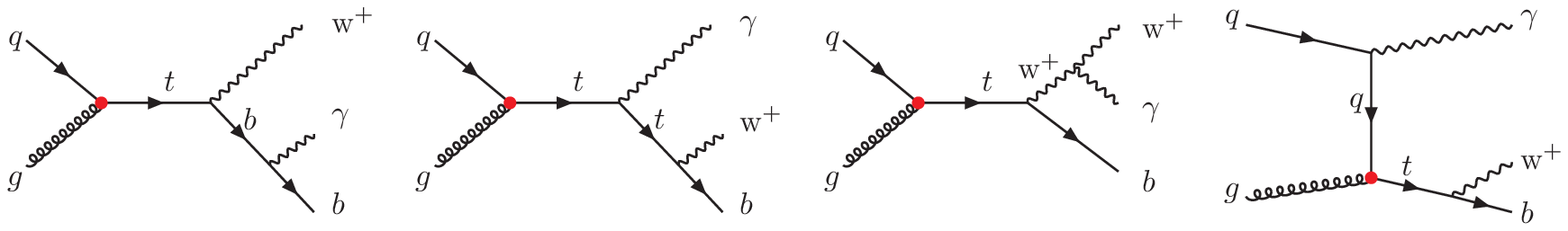}
\caption{  The Feynman diagrams of $p p\to Wb\gamma$ process containing anomalous FCNC $tq\gamma$ (green dot) and $tqg$ (red dot) vertices.\label{fd}}
\end{figure}

\begin{figure}
\includegraphics[scale=0.60]{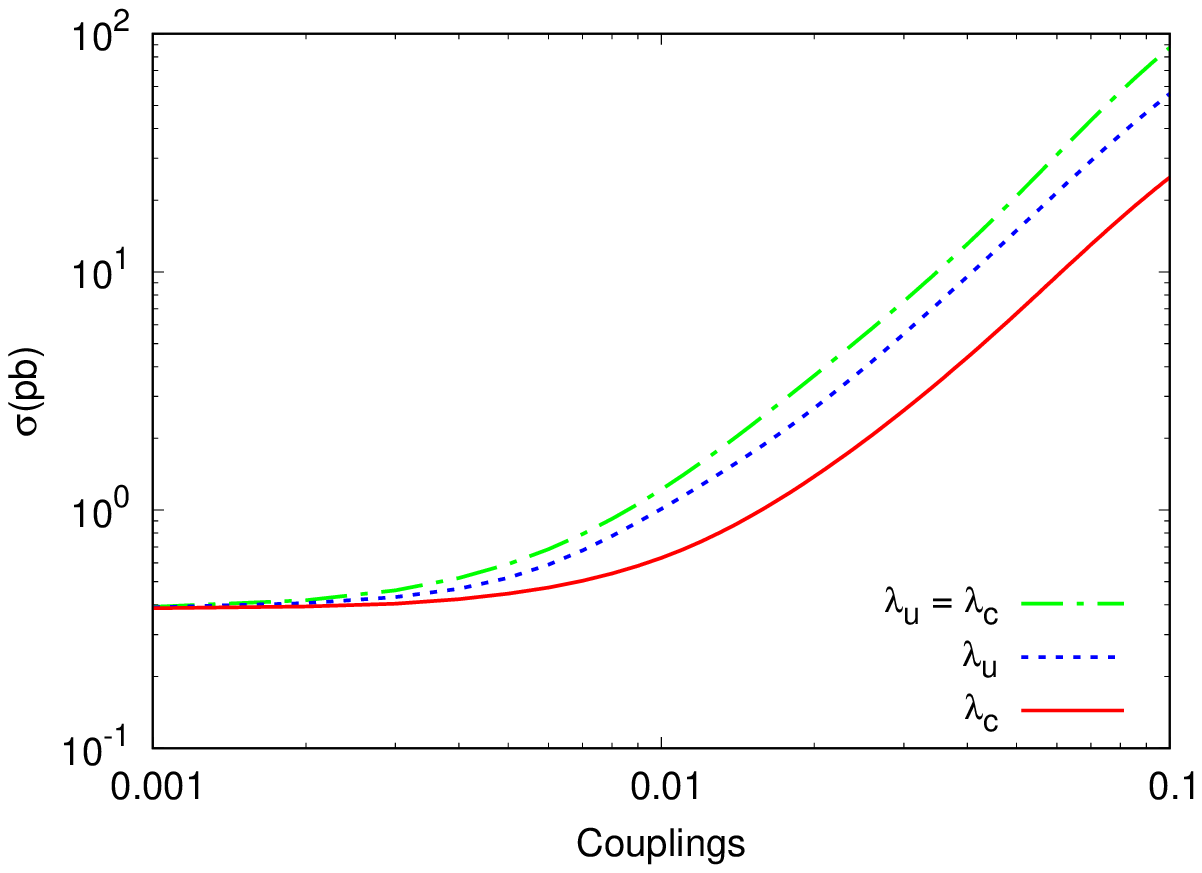} 
\includegraphics[scale=0.60]{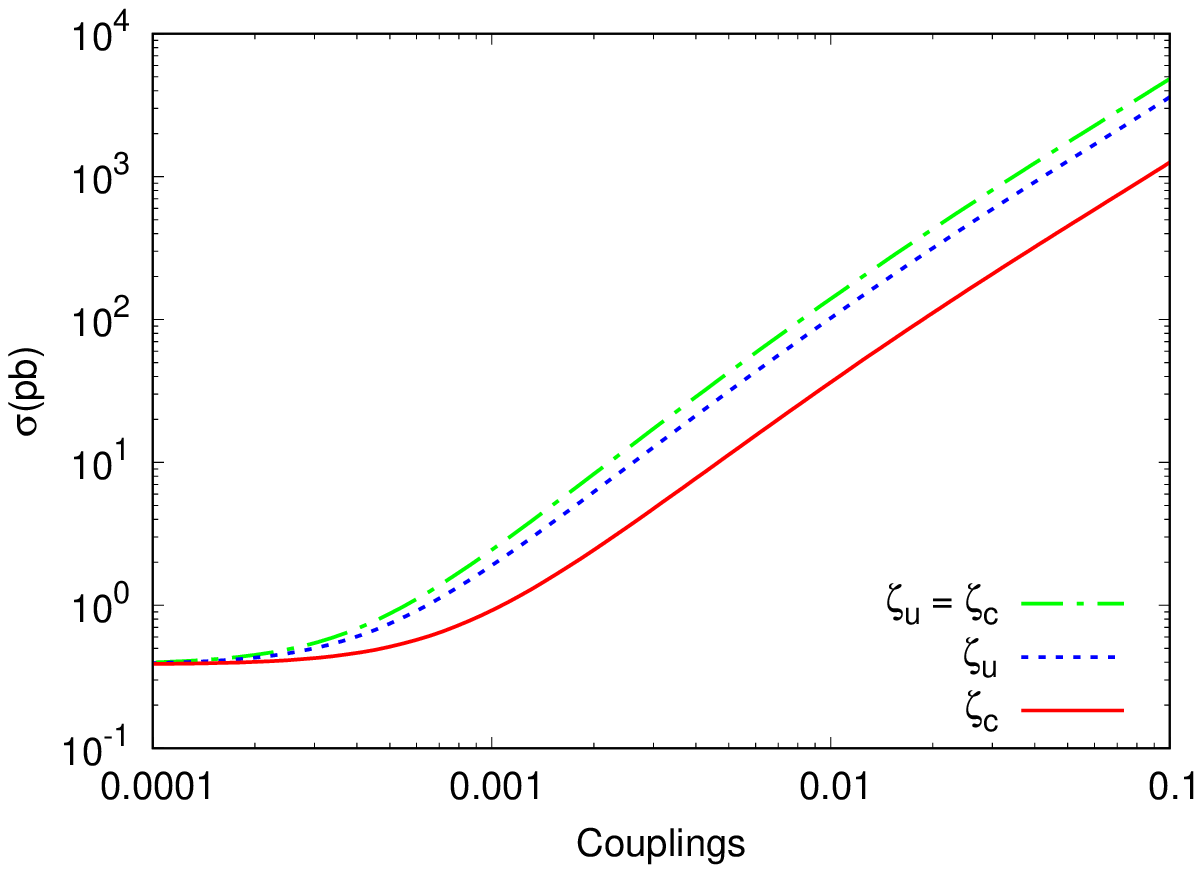}
\caption{  The total cross section of $p p\to Wb\gamma$ process as a function of anomalous FCNC $tq\gamma$ ($\lambda_u$ and $\lambda_c$) and $tqg$ ($\zeta_u$ and 
$\zeta_c$) couplings. \label{cs}}
\end{figure}

In this study, we consider $pp\to W b\gamma$ signal process for searching anomalous FCNC $tqg$ and $tq\gamma$ interactions which denotes in Eq.\ref{eq1}. In the production of signal events, the effective Lagrangian with FCNC couplings is implemented to FeynRules package \cite{Alloul:2013bka} and embedded into \verb|MadGraph2.5.3_aMC@NLO| \cite{Alwall:2014hca} as a Universal FeynRules Output (UFO) module \cite{Degrande:2011ua}. A set of Feynman diagrams contributing to $pp\to W b\gamma$ signal process at tree level are shown in Fig.\ref{fd}. As seen from Fig.\ref{fd}, three diagrams in the first row  contains $tq\gamma$ vertices (green dot) and the four diagrams on the second row contains $tqg$ vertices (red dot). In Fig.2, we show that the total cross sections as a function of $\zeta_q$ and  $\lambda_q$ couplings of $pp\to W b\gamma$ signal process which includes anomalous FCNC $tqg$ and $tq\gamma$ interactions and SM contribution as well as interference between FCNC vertices and SM. As it can be seen from Fig.\ref{cs}, in the region where the value of the couplings is less than 0.005 (0.0005), $tu\gamma$ ($tug$) and $tc\gamma$($tcg$) couplings contribute at the same rate while contribution of $tu\gamma$ ($tug$) is larger  than $tc\gamma$ ($tcg$) coupling for large coupling region since the $up$ quark PDF has the dominant distribution at 100 TeV center of mass energy. In addition, the anomalous contributions are visible for the value of the couplings bigger than 0.005 (0.0005) compared to SM background for $\lambda_q$ ($\zeta_q$). 

Different theoretical frameworks have been used in the literature to describe top quark FCNC in a model independent way. They are based on an effective Lagrangian with D$\geqslant$ 4 operators that satisfy Lorentz and $SU(3)_c \times U(1)_{EM}$ gauge symmetries. The partial wave unitarity and gauge symmetry will be violated at very high energies in an effective theory with large values of anomalous FCNC couplings \cite{Appelquist:1987cf}. Unitarity constraints can set limitations on these couplings for the process $pp \to t \gamma$ as $\lambda_q ( \zeta_q ) < 2m_t/\sqrt {3\alpha_e\alpha_s s / 2}$ which is at the order of $10^{-1}$  with 100 TeV center of mass energy. Satisfying this condition, we performed the analysis for the values of couplings ($\lambda_q$, $\zeta_q$) smaller than 0.1.    

\section{Signal and Background Simulations}
In this section, the analysis of $pp\to W b\gamma$ signal process including the FCNC $tq\gamma$ and $tqg$ couplings as well as relevant SM backgrounds with experimental conditions of the FCC-hh are given. $10^6$ events are generated by \verb|MadGraph2.5.3_aMC@NLO| \cite{Alwall:2014hca} for each signals (using different coupling values) and relevant backgrounds. These generated events are  passed through PYTHIA 8.223 \cite{Sjostrand:2006za} for parton showering and hadronization. The FCC-hh baseline detector configuration embedded into Delphes 3.3.3 via FCC-hh card is used to include detector effects \cite{deFavereau:2013fsa}. During the production of events, produced jets inside the events are clustered by using FastJet 3.2.1 \cite{Cacciari:2011ma} with anti-$k_t$ algorithm where a cone radius is $R$ = 0.4 \cite{Cacciari:2008gp}. Both leptonic ($l\nu$) and hadronic ($jj$) decays of $W$ boson are considered in the analysis of the signal. Then, analysis for $l\nu b\gamma$ and $jjb\gamma$ final states are performed. The relevant background processes and their corresponding cross sections are 
\[\begin{array}{clll}
\bullet&\mathrm{sm}:& pp\to W b \gamma & \sigma=0.38447~\mathrm{pb} \\
\bullet&Wj\gamma: &pp\to Wj' \gamma   &\sigma=1038.3~\mathrm{pb} \\
\bullet&Wj: &pp\to Wj &\sigma=4363~\mathrm{pb} \\  
\bullet&\mathrm{tt}:& pp\to t\bar t &\sigma=25235~\mathrm{pb} \\ 
\bullet&\mathrm{tt} \gamma: &pp\to t\bar t \gamma &\sigma=107.9~\mathrm{pb} \\   
\bullet&Zj \gamma:& pp\to Zj\gamma &\sigma=330.0~\mathrm{pb} \\  
\bullet&\mathrm{jjjj}: &pp\to j'j'j'j'  &\sigma=4.435\times10^8~\mathrm{pb} \\ 
\bullet&\mathrm{bjjj}:& pp\to bj'j'j' &\sigma=1.192\times10^7\mathrm{pb} \\ 
\end{array}
\]
where $j=j', b, \bar b$ and $j'=u,\bar u, d, \bar d, s, \bar s, c, \bar c, g $. The relevant backgrounds sm, $Wj\gamma$, $Wj$, $tt$, $tt\gamma$ and $Zj\gamma$ are considered in $l\nu b\gamma$ final state analysis. In addition to this relevant backgrounds $jjjj$ and $bjjj$ QCD backgrounds are also included in $jjb\gamma$ final states analysis. In order to minimize the effect of experimental issues such as fake photon and mis-tagged b-jet, $Wj\gamma$ and $Wj$ are considered as the other backgrounds since the light jet could be misidentified as b-jet (or photon) candidate. The $tt$ and $tt\gamma$ processes are also added as background events since there are more than one b-jet in the each top decays to $Wb$. The $Zj\gamma$ process is an another SM background in our analysis to include any error in the mass reconstruction of Z and W bosons due to possible inaccuracy of the hadronic calorimeter.

\begin{figure}
\includegraphics[scale=0.4]{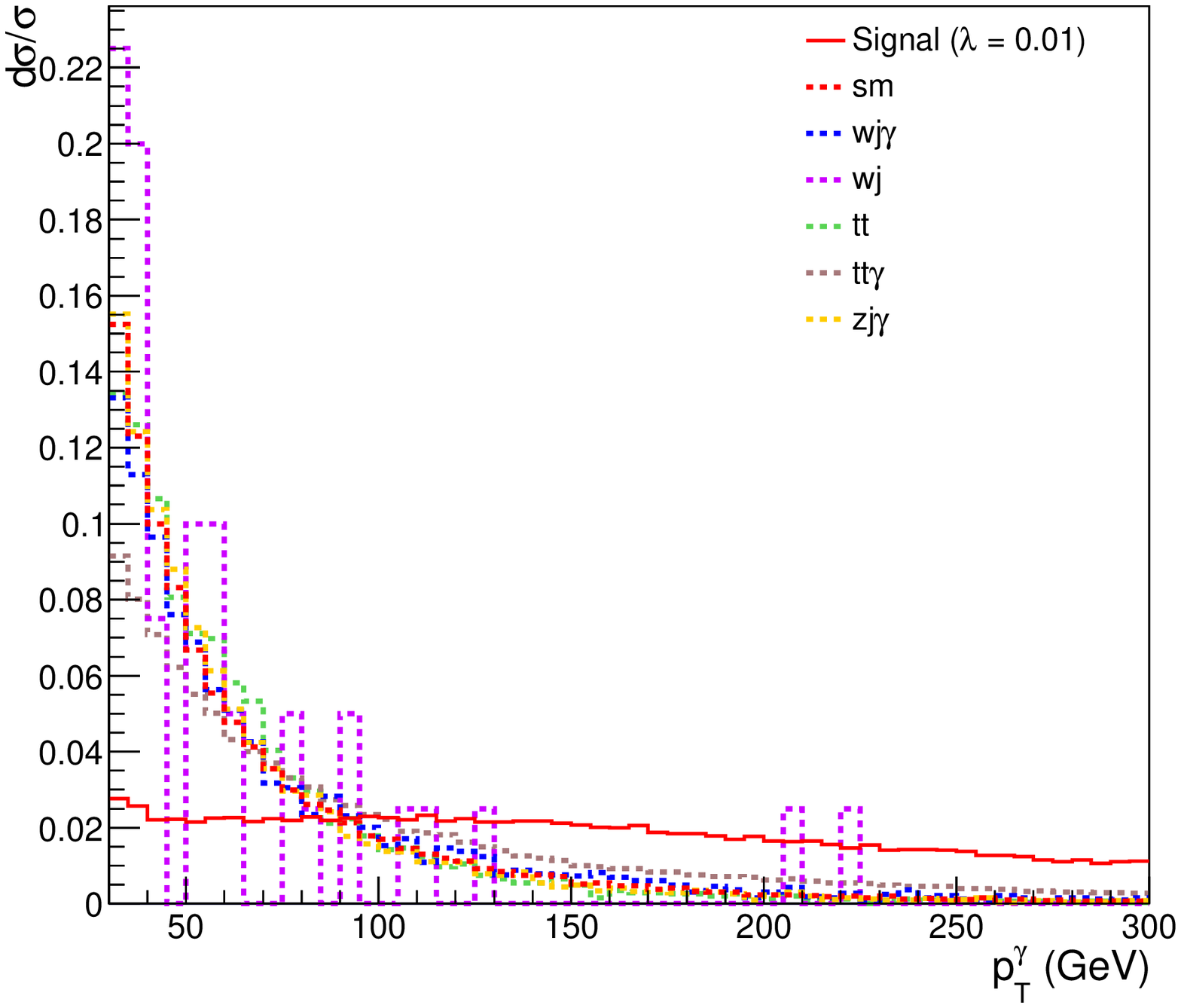} 
\includegraphics[scale=0.4]{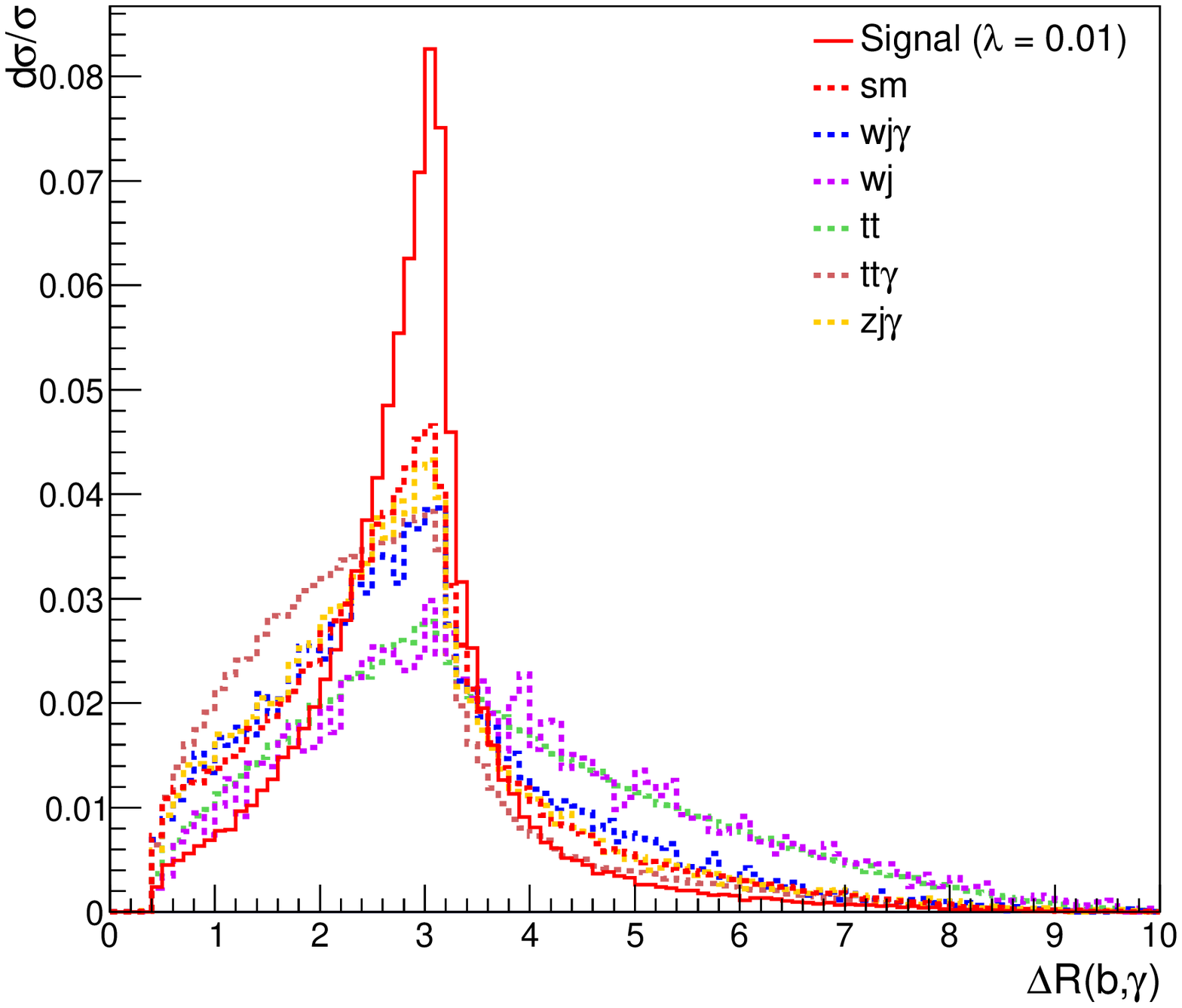}\\ 
\includegraphics[scale=0.4]{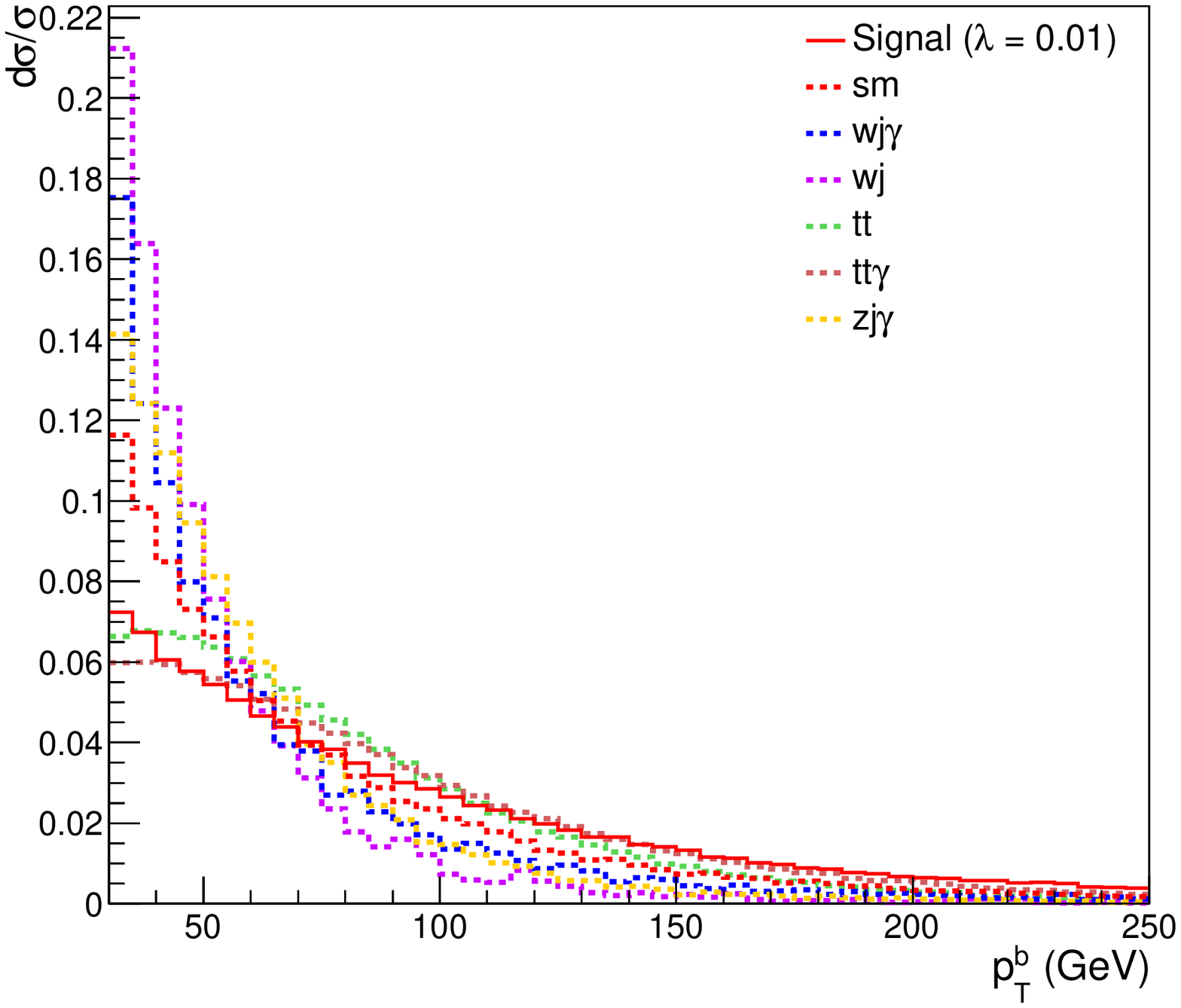} 
\includegraphics[scale=0.4]{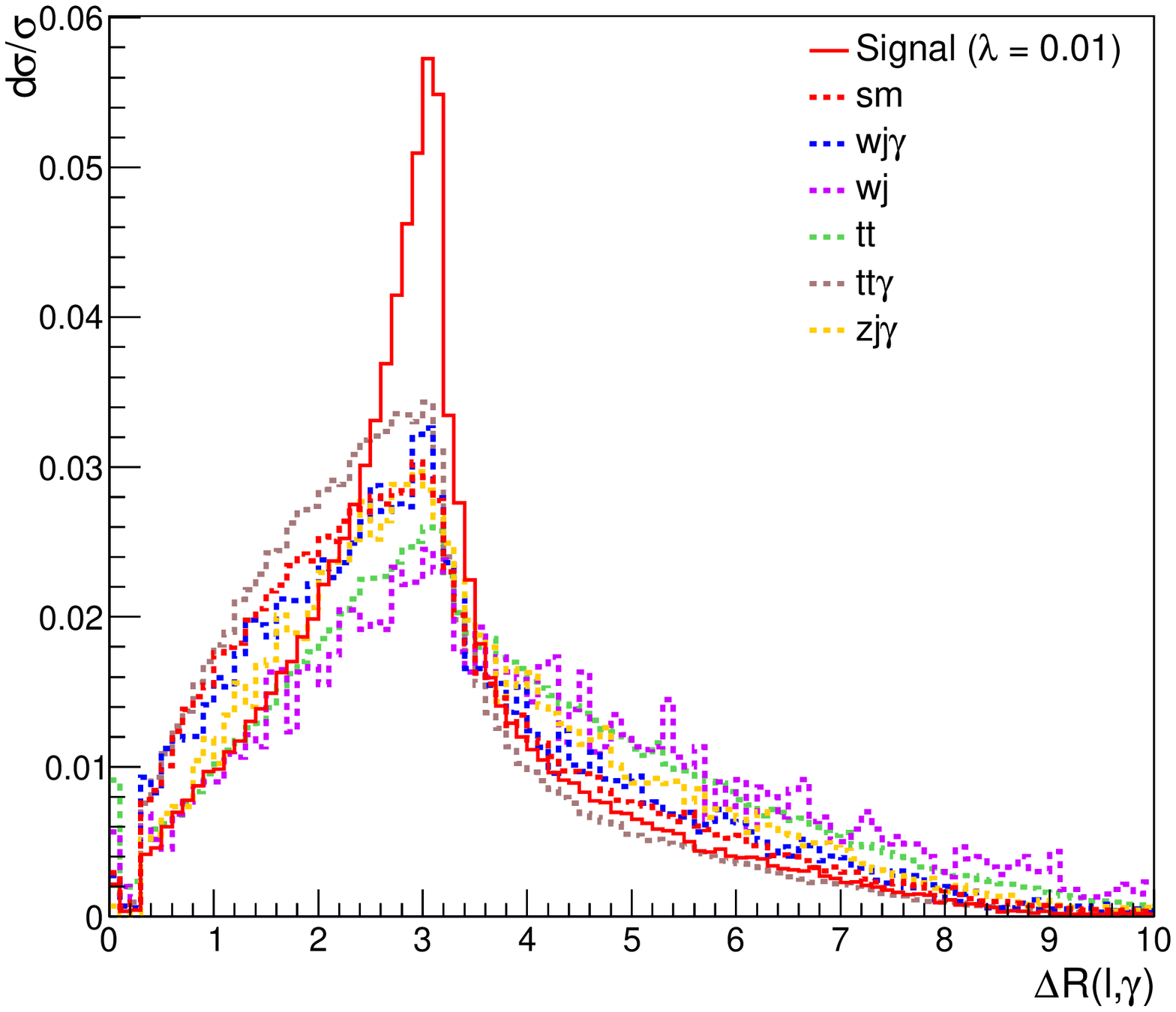}\\ 
\includegraphics[scale=0.4]{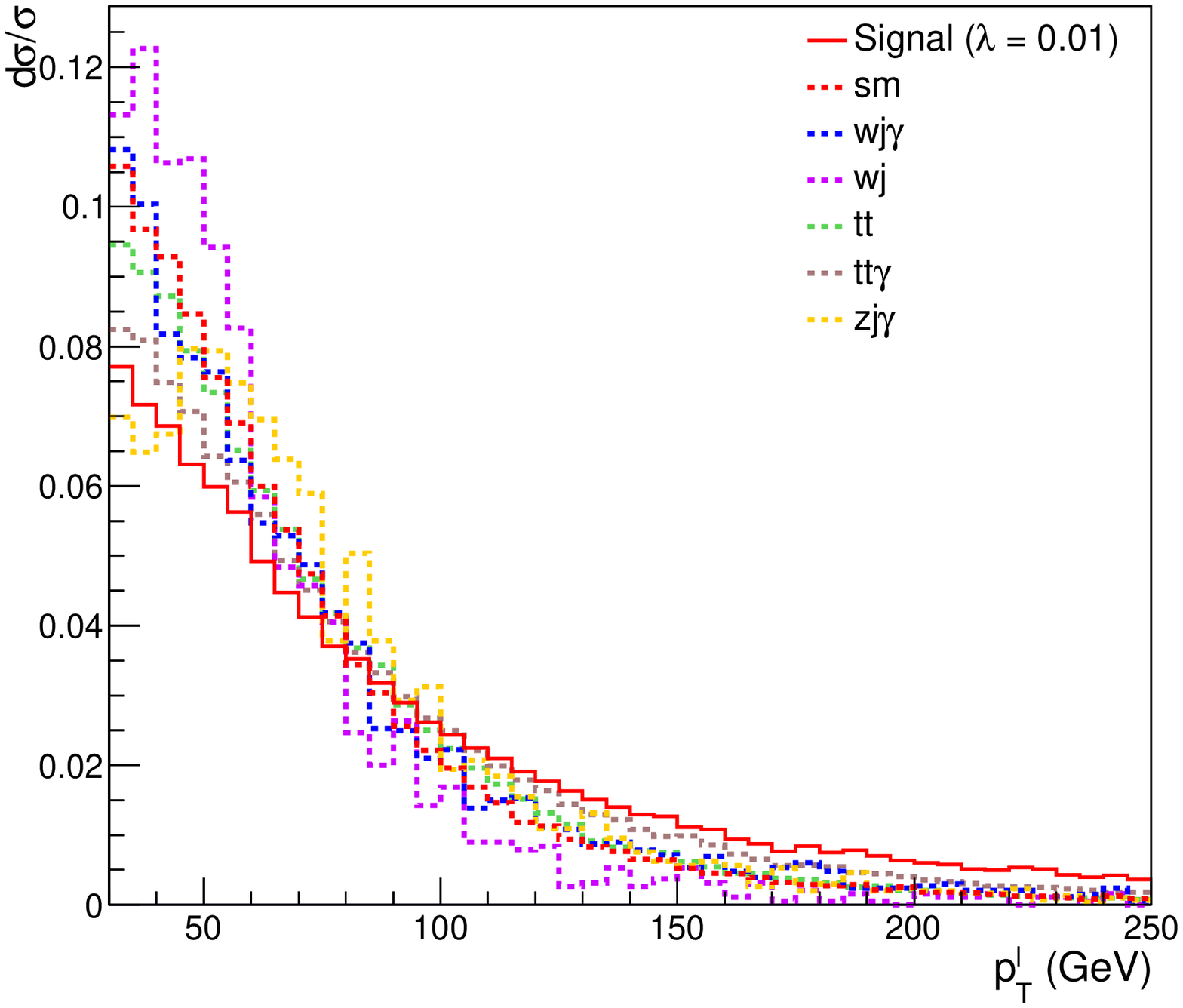}
\includegraphics[scale=0.4]{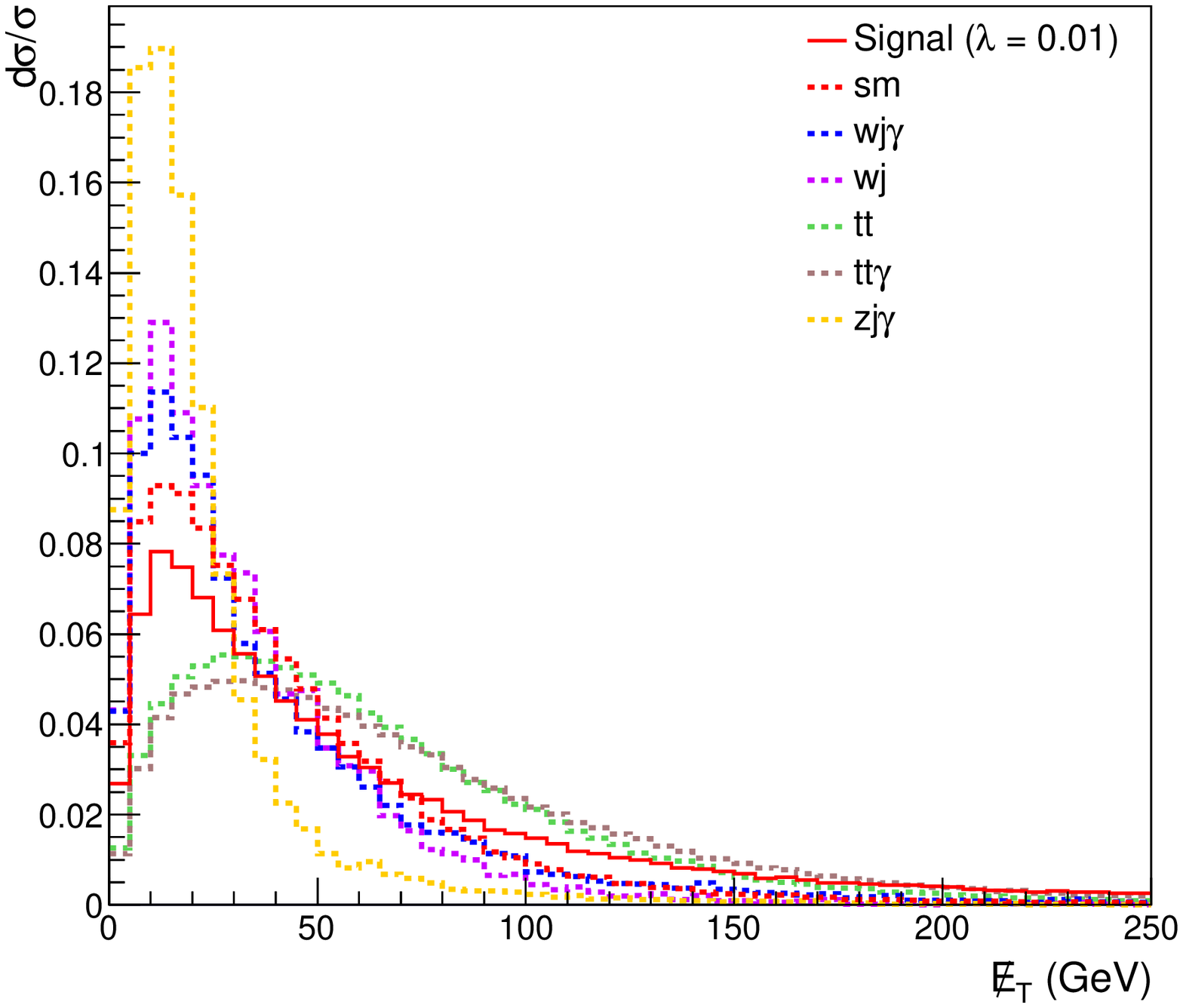}
\caption{The kinematic distributions of the final state particles in leptonic channel for signal ($\lambda_q = 0.01$) and relevant SM background processes; transverse momentum of  $\gamma$, $b$-jet and lepton on the left column and $\Delta R (b,\gamma)$, $\Delta R (l,\gamma)$ and MET on the right column.  \label{kin_lep}}
\end{figure}

\begin{figure}
\includegraphics[scale=0.4]{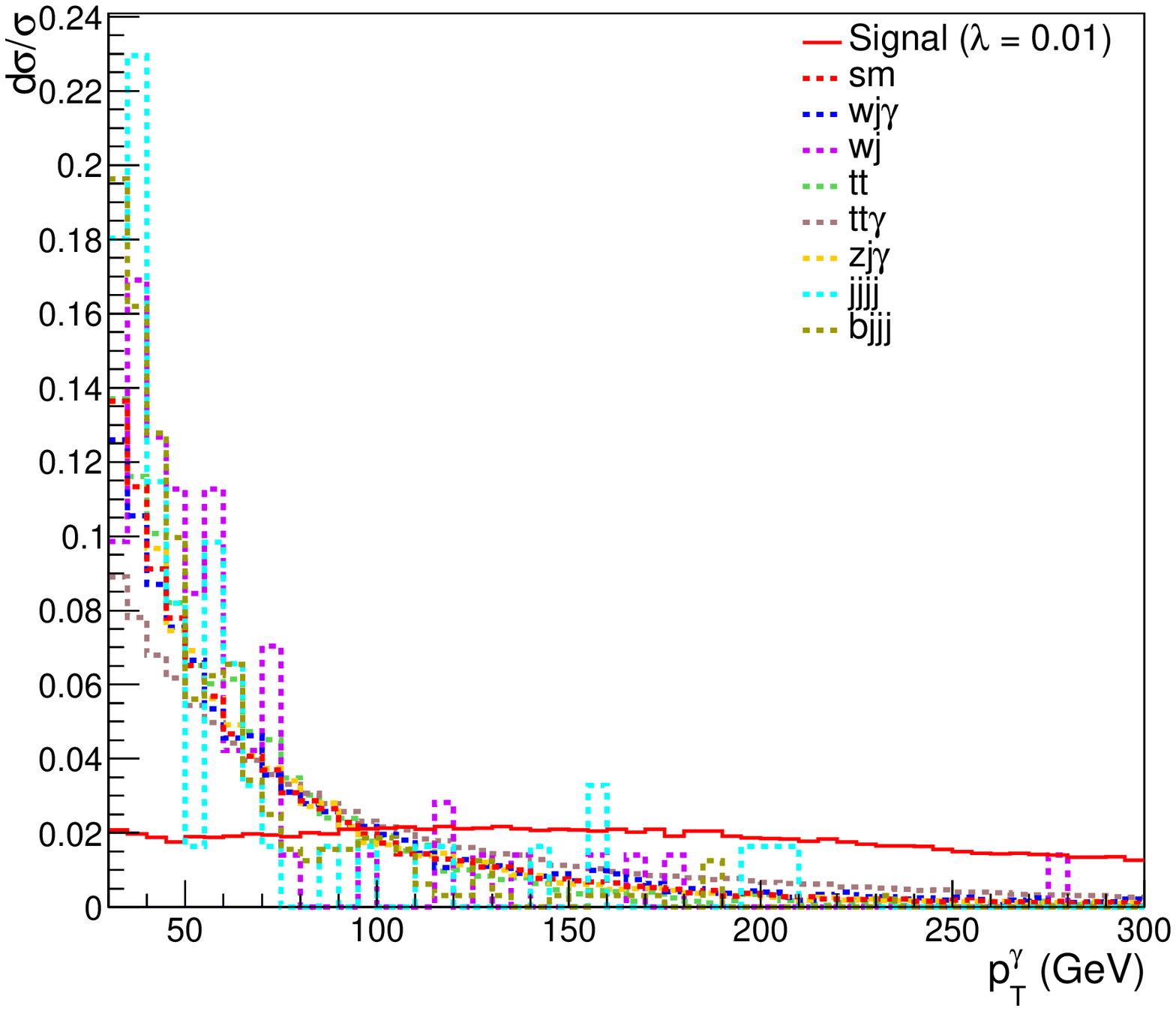} 
\includegraphics[scale=0.4]{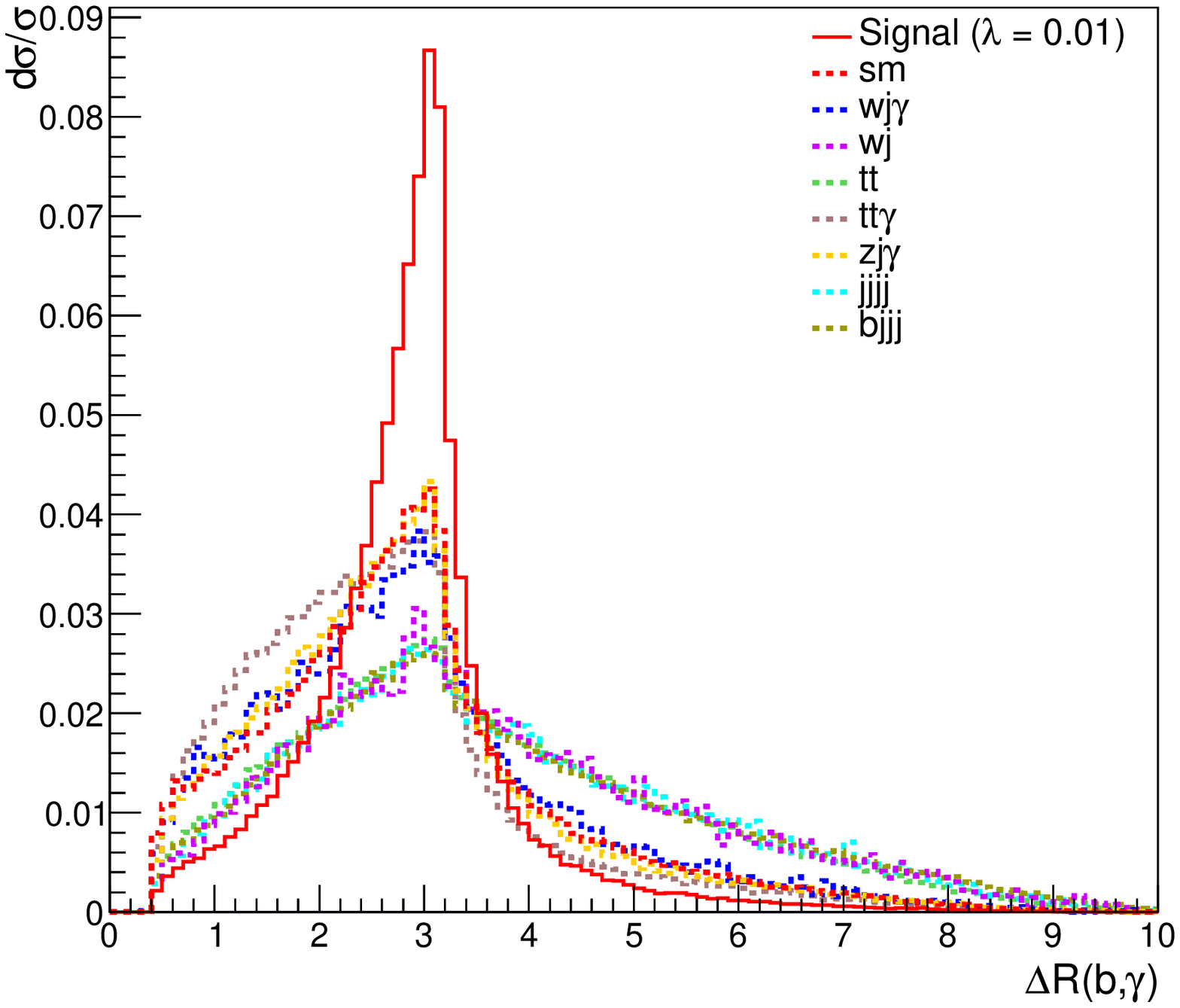}\\ 
\includegraphics[scale=0.4]{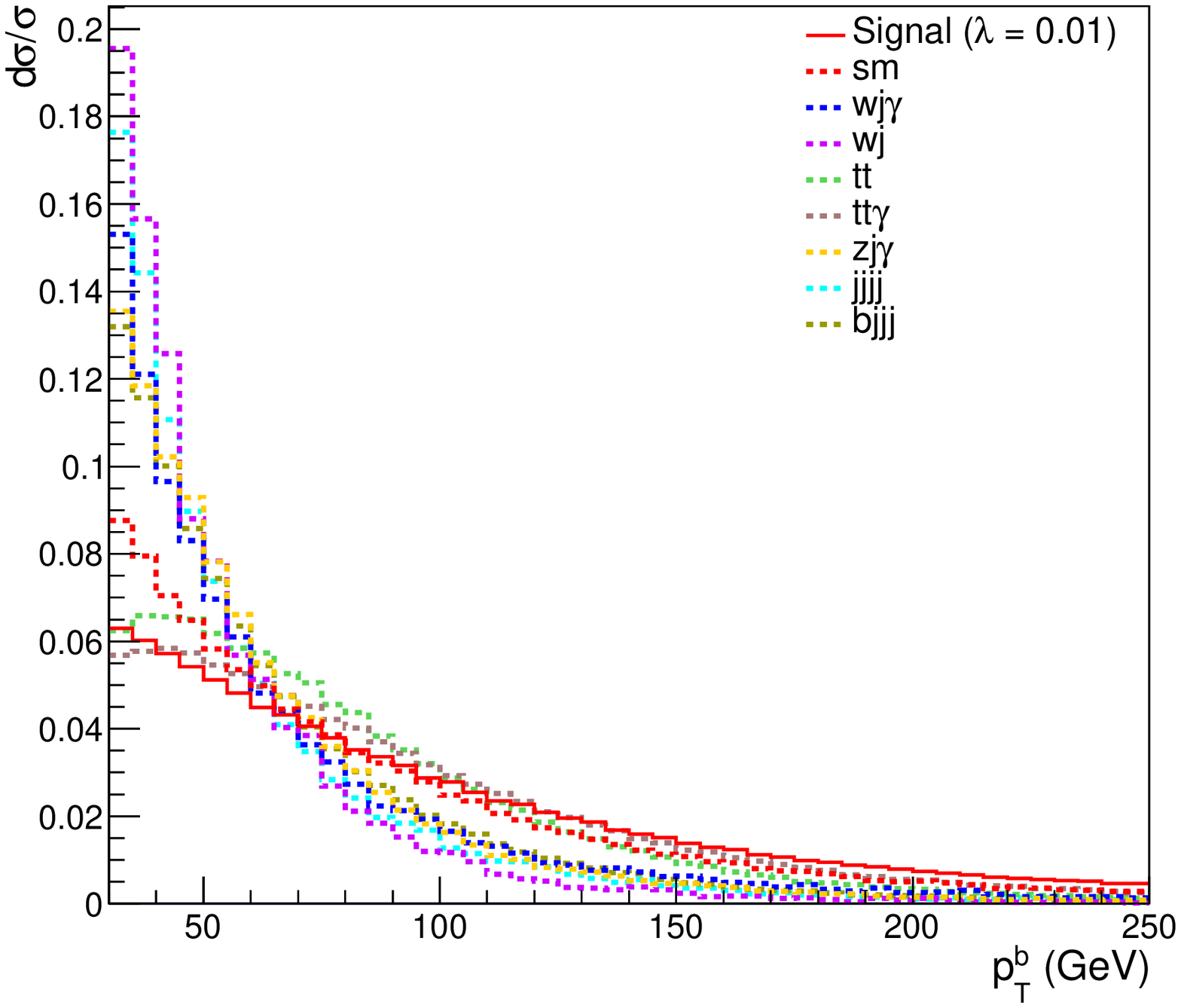} 
\includegraphics[scale=0.4]{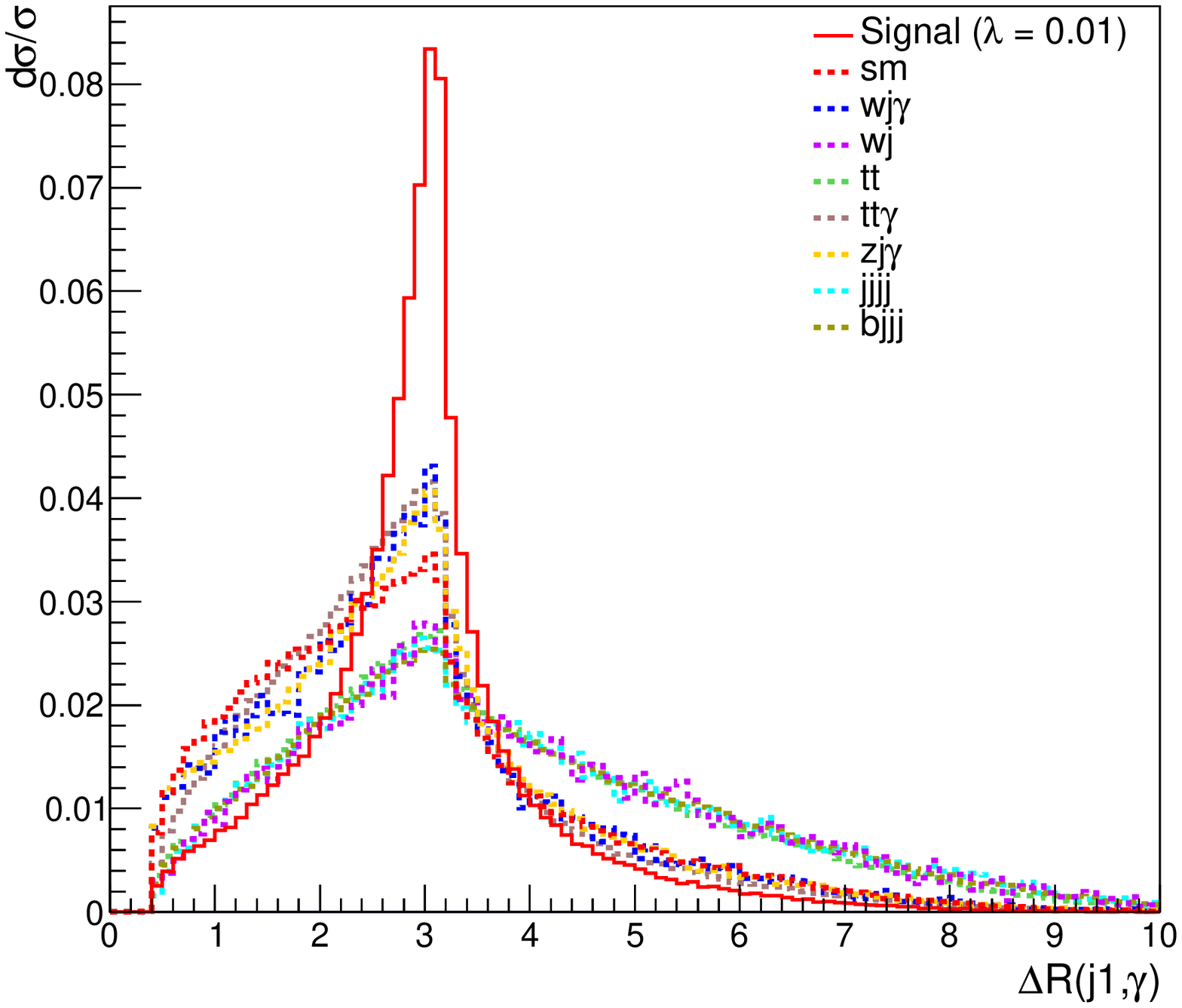}\\ 
\includegraphics[scale=0.4]{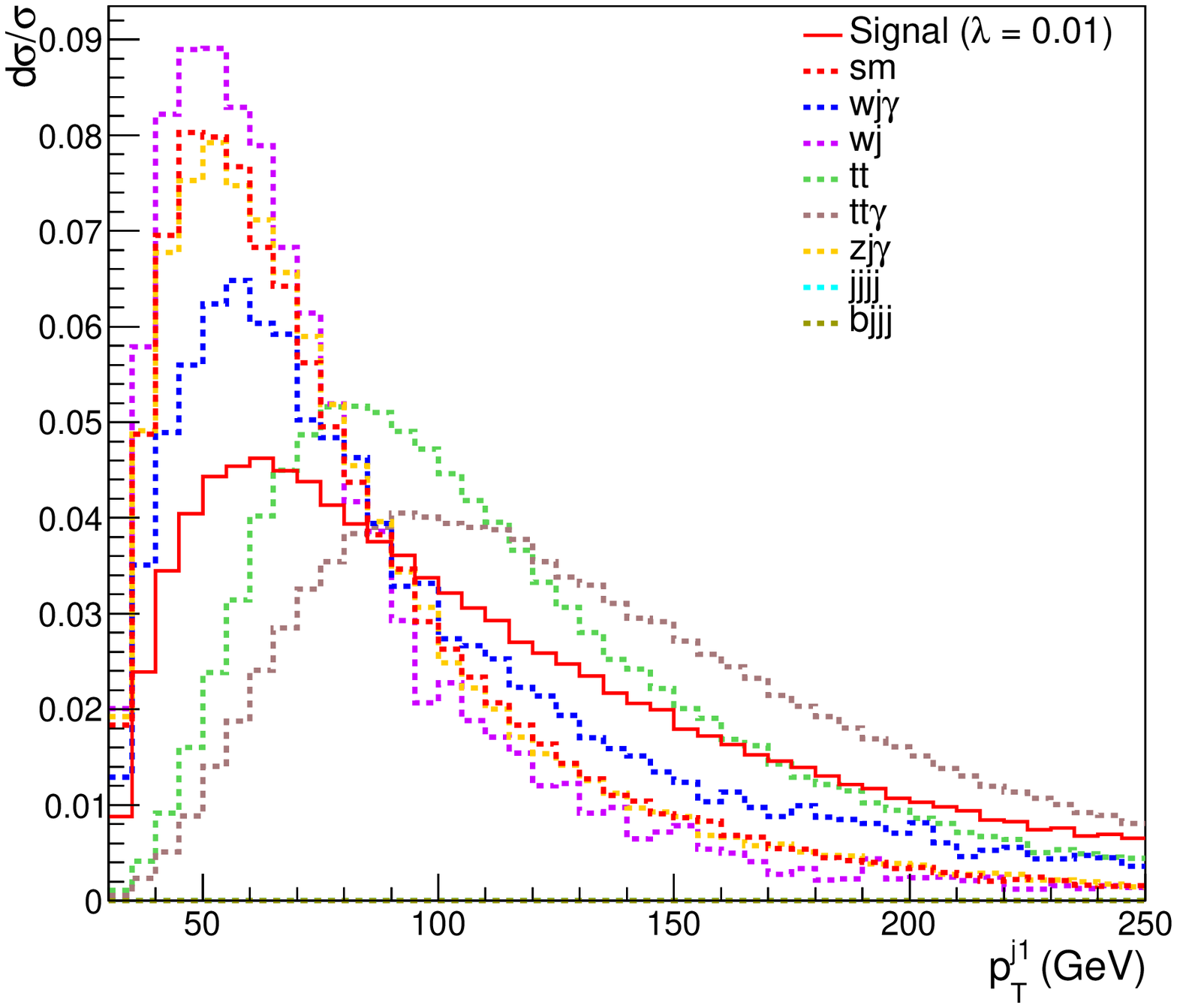}
\includegraphics[scale=0.4]{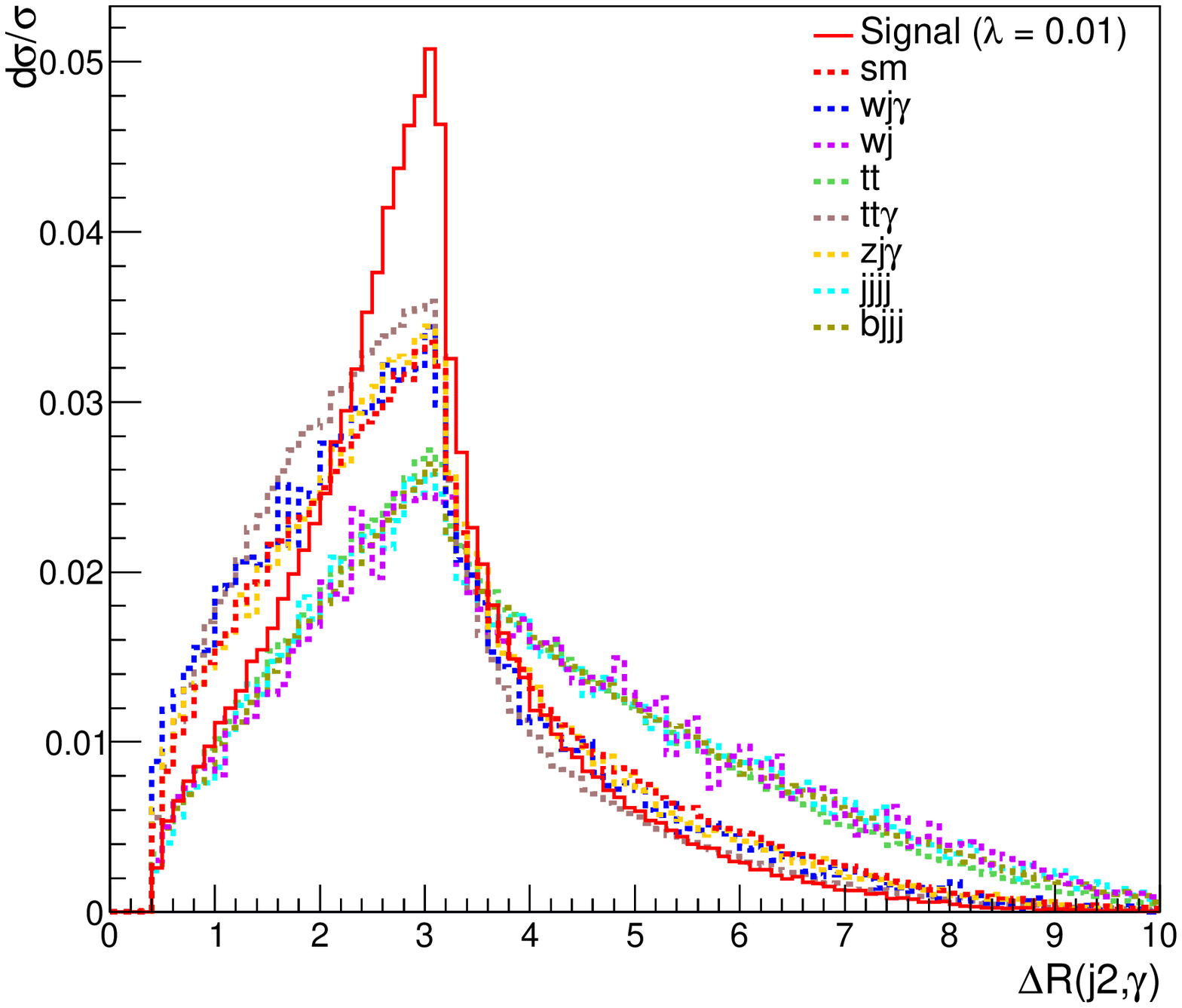}
\caption{The kinematic distributions of the final state particles in hadronic channel for signal ($\lambda_q = 0.01$) and relevant SM background processes; transverse momentum of  $\gamma$, $b$-jet and leading jet ($j1$) on the left column and $\Delta R (b,\gamma)$, $\Delta R (j1,\gamma)$ and $\Delta R (j2,\gamma)$ on the right column.  \label{kin_had}}
\end{figure}

\begin{figure}
\includegraphics[scale=0.4]{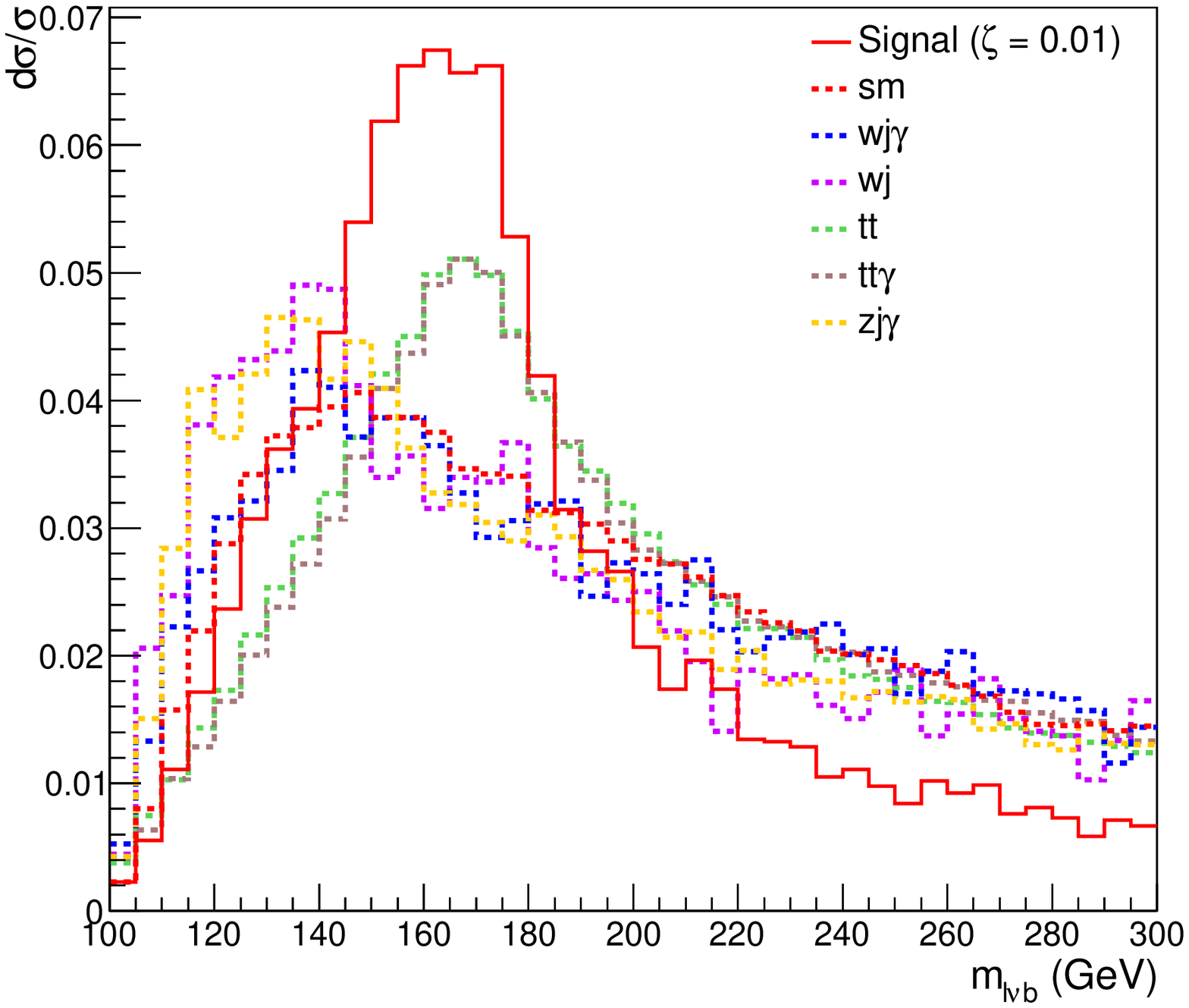} 
\includegraphics[scale=0.4]{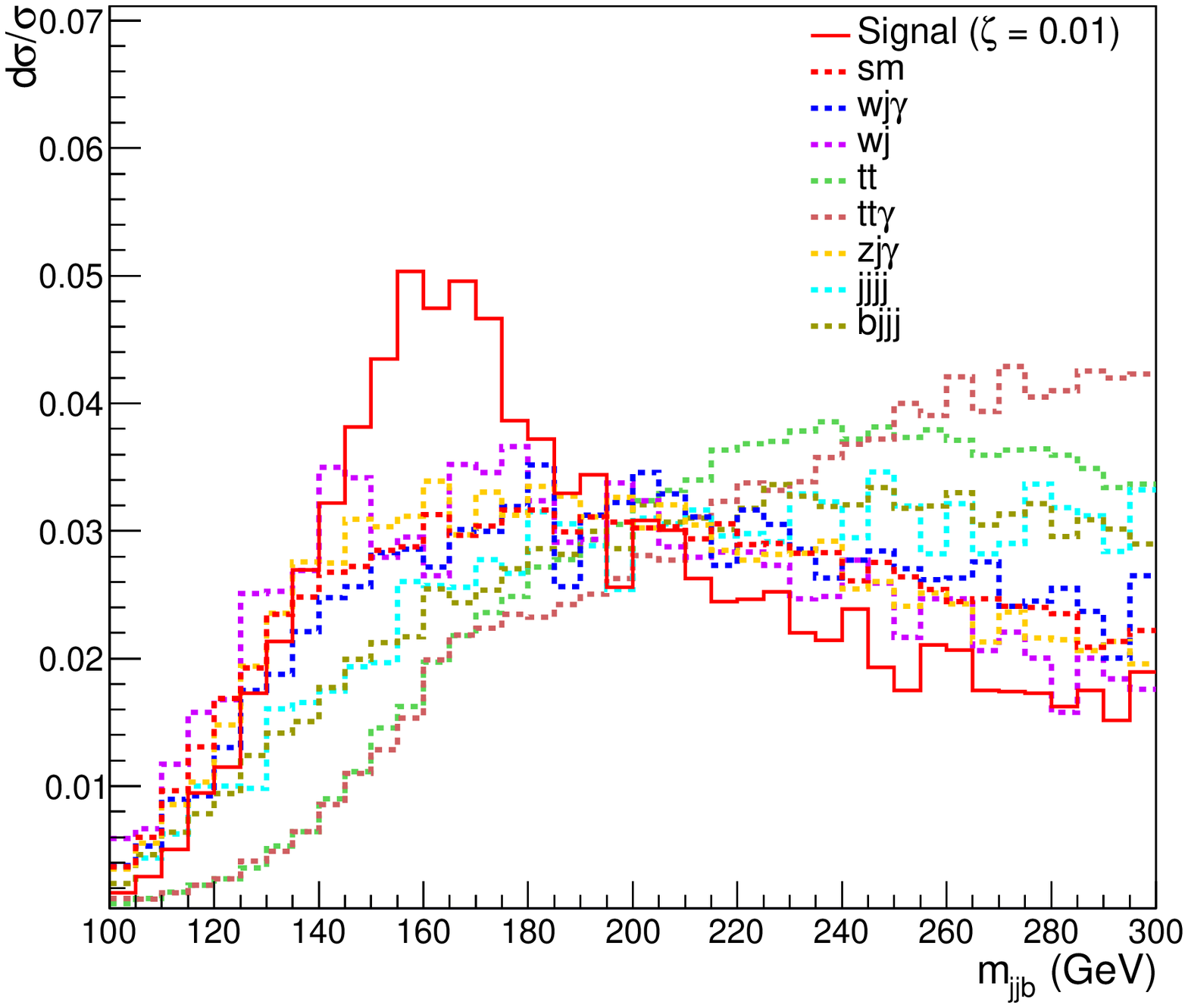}\\
\includegraphics[scale=0.4]{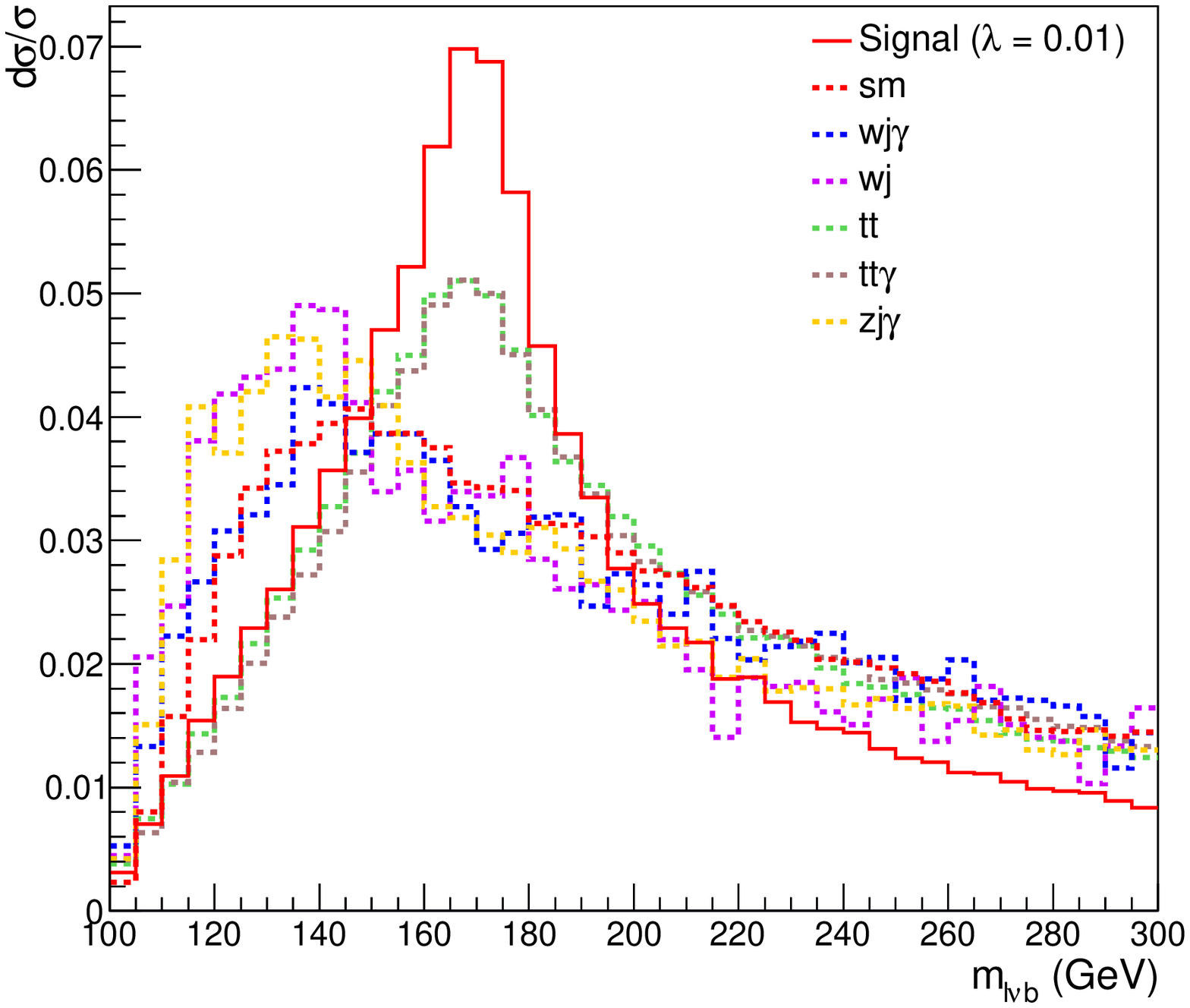} 
\includegraphics[scale=0.4]{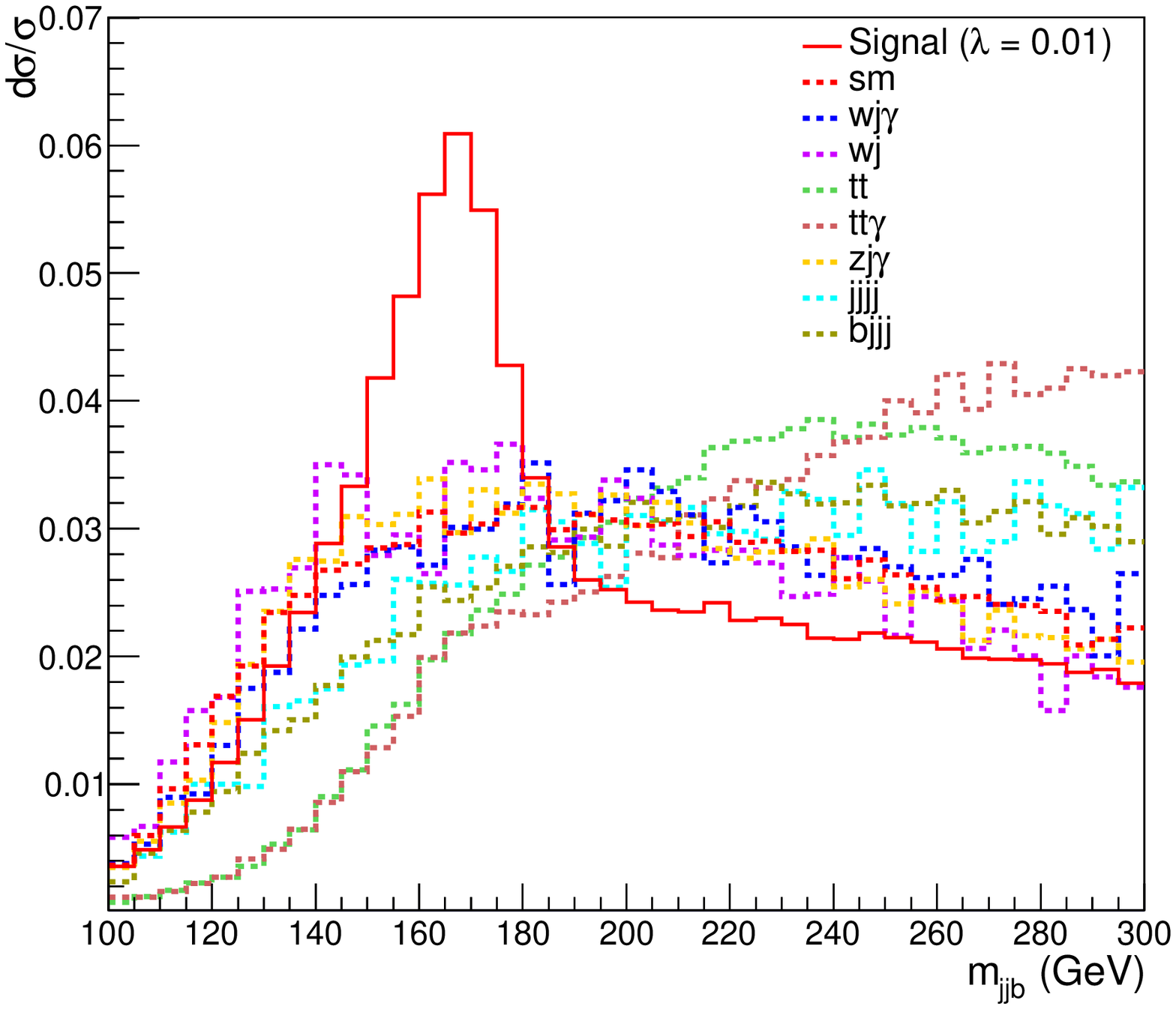}
\caption{The reconstructed invariant mass distributions of signal ($\zeta_q = 0.01$ and $\lambda_q = 0.01$ on the top and bottom, respectively) and relevant SM background processes for leptonic (on the left) and  hadronic channel (on the right).  \label{minv_had_lep}}
\end{figure}

In order to distinguish signal from relative backgrounds, different preselection and kinematical cuts are applied separately to hadronic and leptonic channels of $W$ boson in the signal process as follows:
In the leptonic channel of signal, at least one photon ($N_\gamma \geqslant1$) and one lepton ($N_l \geqslant 1$) are required with one isolated b-jet ($N_b = 1$) as a preselection cut. On the other hand, at least one photon ($N_\gamma \geqslant 1$ ) and three jets ($N_j \geqslant 3$), one of them is isolated b-jet ($N_b = 1$), with no lepton are applied as a preselection cut in the hadronic channel of the signal. By these preselection cuts, not only b-jet rich backgrounds but also multijet backgrounds that contain mis-tagged particles in their events are eliminated for effective analysis of hadronic and leptonic signal channels. Kinematic distributions of the final state particles for leptonic and hadronic channels after pre-selection are given in Fig.\ref{kin_lep} and Fig.\ref{kin_had}, respectively. In Fig.\ref{kin_lep}, the Missing Energy Transverse (MET) distribution,  a separation between a photon and b-jet $\Delta R (b,\gamma)$ as well as photon and lepton $\Delta R (l,\gamma)$ in the pseudorapidity-azimuthal angle plane, the transverse momentum distributions of the photon, lepton, and b-jet are shown. In Fig.\ref{kin_had}, the transverse momentum distributions of the photon, first leading jet ($j1$), b-jet,  a separation between a photon and b-jet $\Delta R (b,\gamma)$, photon and $j1$ $\Delta R (j1,\gamma)$  as well as  photon and second leading jet ($j2$) $\Delta R (j2,\gamma)$  in the pseudorapidity-azimuthal angle plane are depicted. Firstly, we applied cut on the transverse momentum of leading photon 50 GeV as well as other kinematical cuts for the final state particles (kinematic-I). As seen from Fig.\ref{kin_lep} and Fig.\ref{kin_had}, photons have large momentum because of recoil against the heavy top quark.  Thus, leading photon with $p_T^{\gamma}$ > 150 GeV (kinematic-II) is required to distinguish signal from backgrounds in both channels as well as  other optimal kinematical cuts summarized in Table \ref{tab1}. Two leading light jets are used to reconstruct $W$ boson for hadronic channel while lepton and neutrino for leptonic channel. Since four-momentums of the leading and second-leading jets are  precisely measured, one can reconstruct $W$ mass easily for hadronic channel. However, for the reconstruction of $W$-boson in leptonic channel, one needs to know the longitudinal component of the neutrino momentum ($p_{z,\nu}$) has to be taken into account. The $p_{z,\nu}$ is obtained by missing transverse energy of the neutrino ($\slashed {E}_T$) and energy-momentum conservation in the $Wl\nu$ vertex:
\begin{eqnarray}
p_{z,\nu}^{\pm}&=&\frac{1}{p_{T,l}^2} \Big( \Lambda p_{z,l}\pm \sqrt{\Lambda^2p_{z,l}^2-p_{T,l}^2(E_l^2\slashed {E}_T^2-\Lambda^2)} \Big)
\end{eqnarray}
where $\Lambda=(M_W^2/2)+\vec{p}_{T,l}\cdot \slashed{\vec p}_T$; the $E_l$, $p_{T,l}$ and $p_{z,l}$ are the energy, transverse and longitudinal momentum components of the leading lepton, respectively.  We chose the solution with the smallest absolute value of $p_{z,\nu}$  because the true $p_{z,\nu}$ is about 70\% \cite{Belyaev:1998dn}. 
For both leptonic and hadronic channels, constraints on mass range of the reconstructed $W$ boson as well as the reconstructed top quark which is the vector sum of the 4-momenta of reconstructed $W$-boson and $b$-tagged jet  are used as in Table \ref{tab1}. Effects of cuts defined in Table \ref{tab1} on the number of events with $L_{int}=100$ fb$^{-1}$ can be seen in Table \ref{tab2} and Table \ref{tab3}  for leptonic and hadronic channels, respectively. Specially kinematic-I cut set reduces $Wj$, $tt$ and $Zj\gamma$ backgrounds while selecting high $p_T^{\gamma}$ cut (kinematic-II) effects other backgrounds as well. For example, the cut efficiency of kinematic-I after pre-selection is about 28.5 \% for signal ($\lambda=0.01$), 4.1 \% for sm background which has the same final state with signal, 3.6 \% for $Wj\gamma$,  0.02 \% for $Wj$, 0.23 \% for tt , 13 \% for tt$\gamma$ and  0.44 \% for $Zj\gamma$ in the leptonic channel. Applying kinematic-II cut enhance the cut efficiency further one order. In Fig.\ref{minv_had_lep}, the reconstructed invariant mass distributions of signal ($\zeta_q = 0.01$ and $\lambda_q = 0.01$ on the top and bottom, respectively) and relevant SM background processes for leptonic (on the left) and hadronic channel (on the right) are plotted in the mass window with the pre-selection cut. The sharp signal peaks for both leptonic and hadronic channels are clearly seen above broad relevant backgrounds in the invariant mass distributions. Therefore, we require reconstructed invariant mass window between 135 GeV and 195 GeV to calculate Statistical Significance (SS).

\begin{figure}
\includegraphics[scale=0.6]{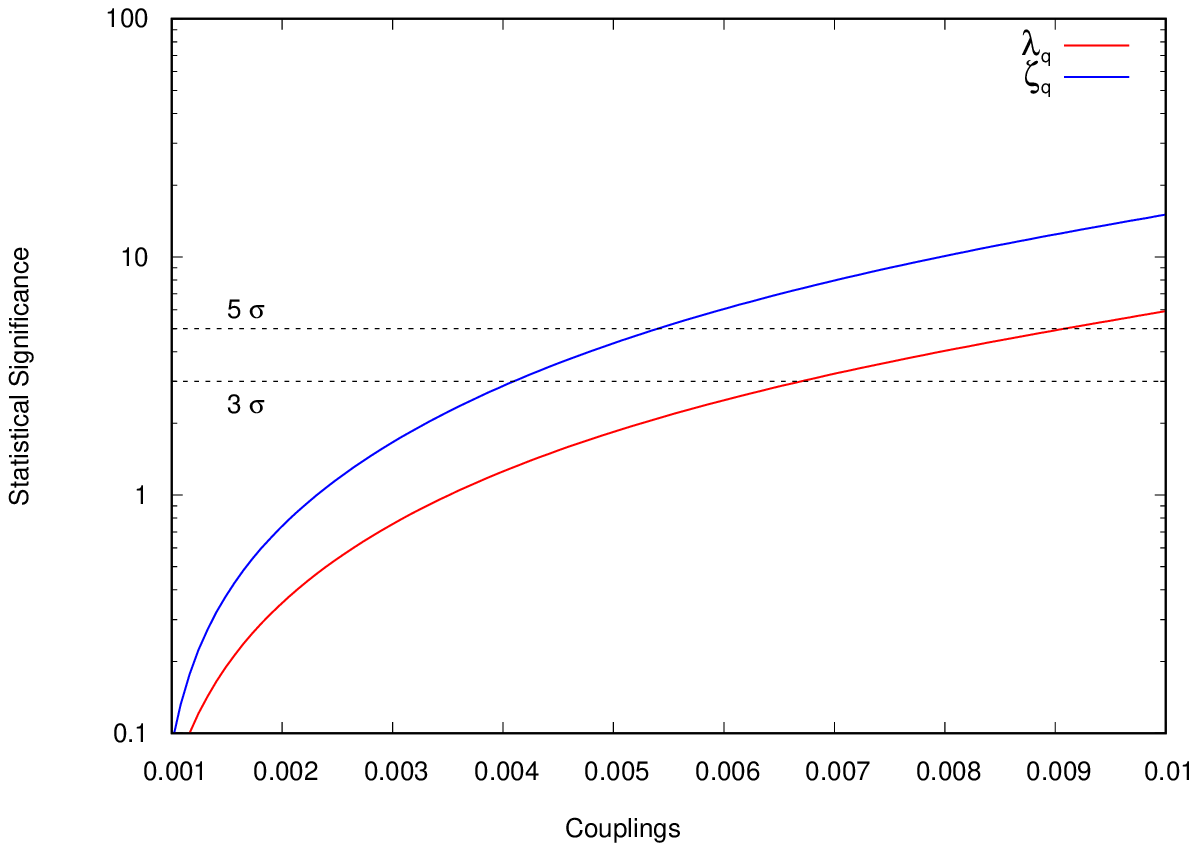} 
\includegraphics[scale=0.6]{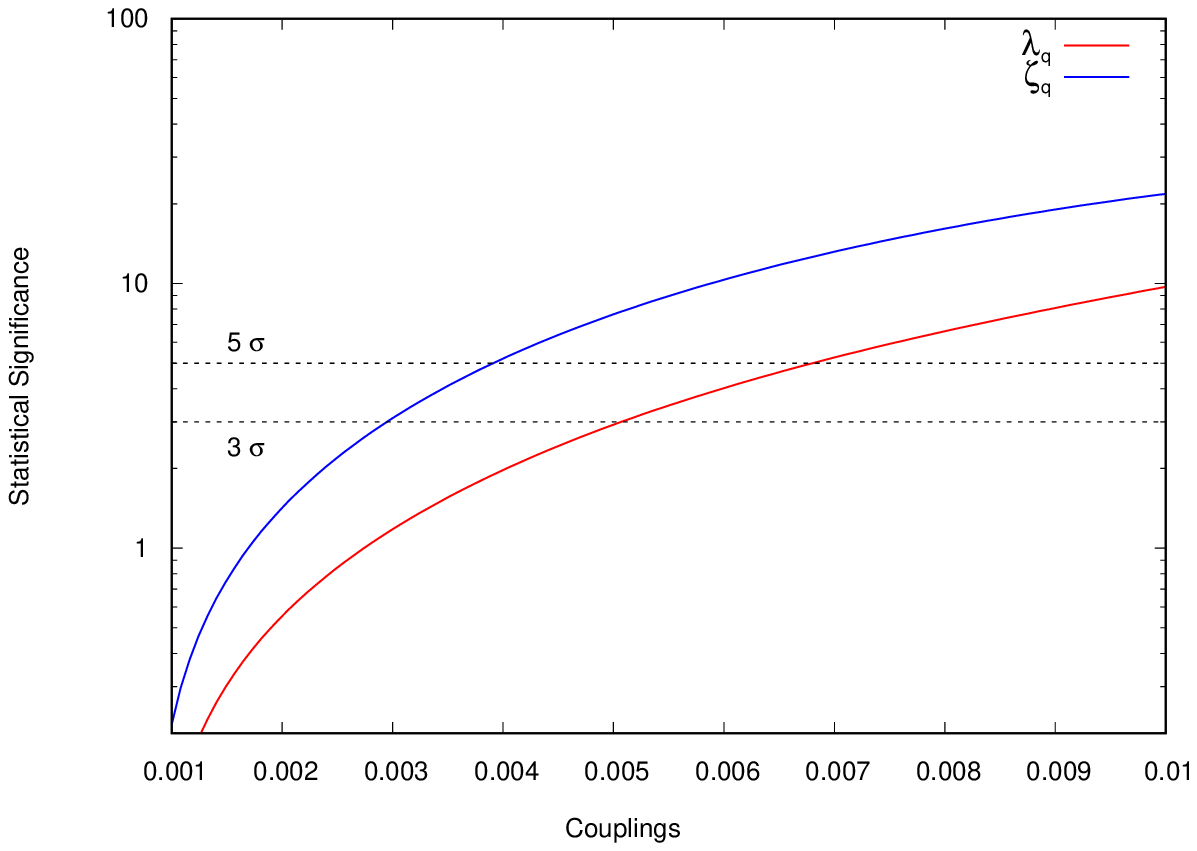}
\caption{The statistical significance as a function of the anomalous FCNC top couplings strengths after applying all cuts for leptonic (on the left) and hadronic (on the right)  channels at L$_{int}$=100 fb$^{-1}$. Only one coupling ($\lambda_q$ or $\zeta_q$) at a time is varied from its SM value).  \label{ss_one}}
\end{figure}

\begin{figure}
\includegraphics[scale=0.45]{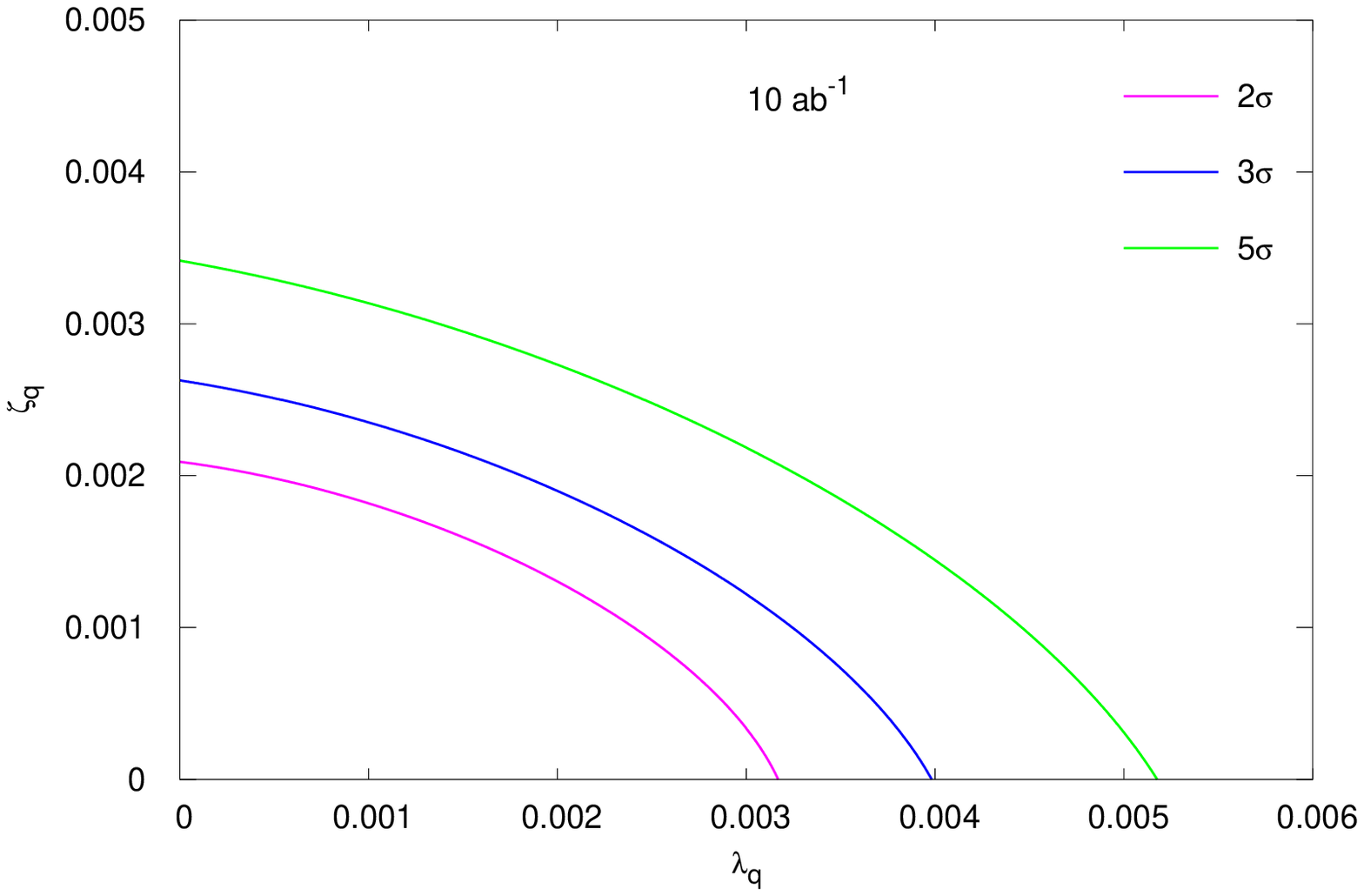} 
\includegraphics[scale=0.45]{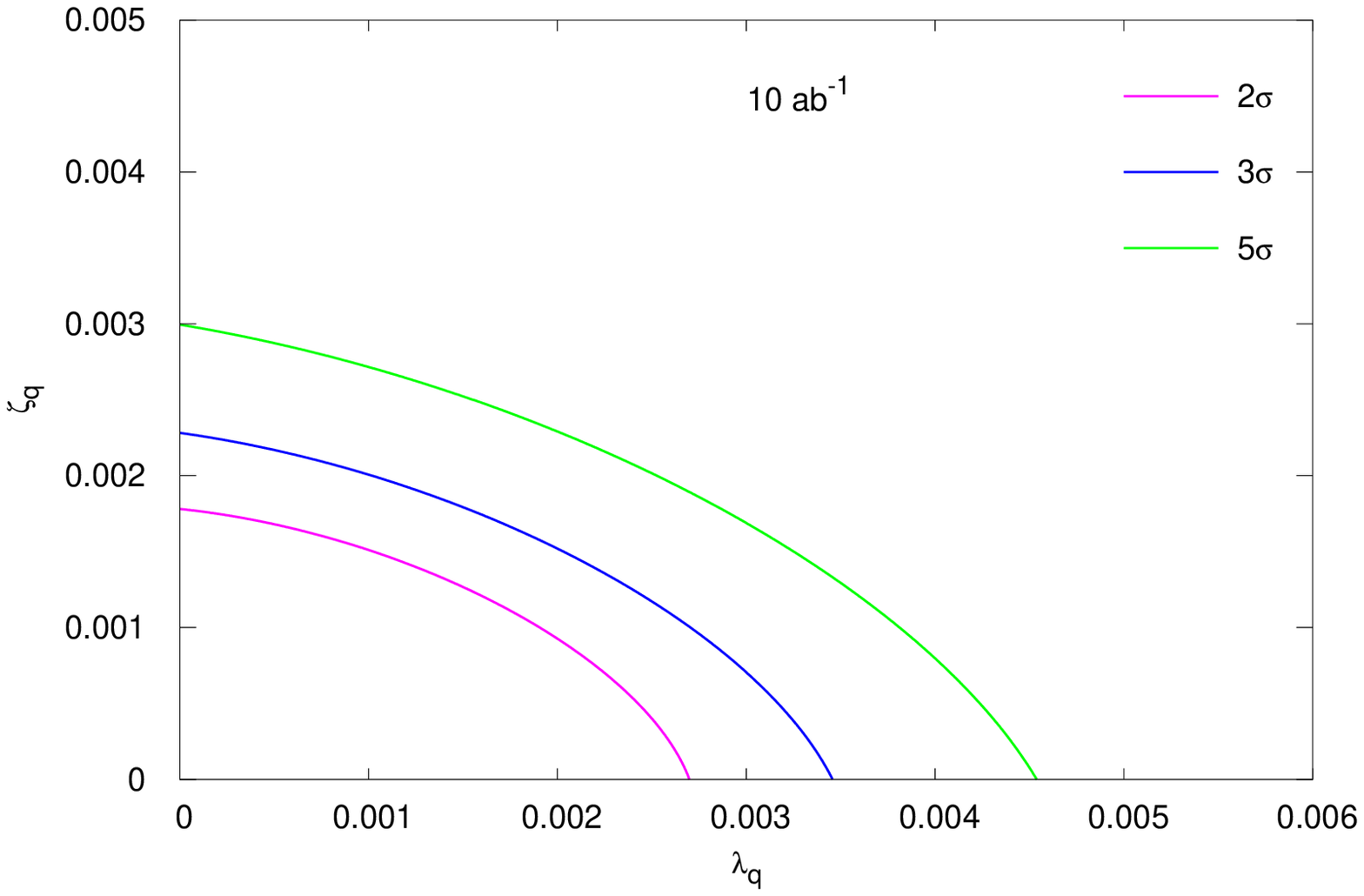}
\caption{The contour plots of $2\sigma$, $3\sigma$ and $5\sigma$ significance on the $\lambda_q$-$\zeta_q$ anomalous FCNC couplings plane with an integrated luminosity of 10 ab$^{-1}$ for leptonic (on the left) and hadronic (on the right) channels.  \label{ss_two}}
\end{figure}

Using Poisson formula
\begin{equation}
SS=\sqrt{2[(S+B_T)\ln(1+S/B_T)-S]}
\end{equation}
where $S$ and $B_T$ are the signal and total background events at a particular luminosity. The results for the $SS$ values depending on the couplings $\lambda_q$ and $\zeta_q$ at L$_{int}$=100 fb$^{-1}$  for leptonic (on the left) and hadronic (on the right) are given in Fig.~\ref{ss_one}.  In this figure, only one coupling ($\lambda_q$ or $\zeta_q$) at a time is varied from its SM value and $3\sigma$ and $5\sigma$ discovery ranges are presented. It is clear from Fig.~\ref{ss_one} that the FCC-hh would reach $\lambda_q$=0.0065 (0.005) while $\zeta_q$=0.0041 (0.0028) at $3\sigma$ significance for leptonic (hadronic) channel. We also simultaneously vary both anomalous top couplings to find excluded region in  $\lambda_q$-$\zeta_q$ plane.  The boundary of $2\sigma$, $3\sigma$ and $5\sigma$ excluded region in $\lambda_q$-$\zeta_q$ plane for leptonic (on the left) and hadronic (on the right) channels with an integrated luminosity 10 ab$^{-1}$ at 100 TeV are plotted in Fig.~\ref{ss_two}. For both anomalous top couplings at $5\sigma$ with L$_{int}$=100 fb$^{-1}$ gives better results than at $3\sigma$  with L$_{int}$=100 fb$^{-1}$ as seen in Fig.~\ref{ss_two}.  

One can express results in terms of branching ratios which can be comparable with the results
of other studies. Both FCNC decay widths and total decay width ($\Gamma(t \to Wb$)) of the top quark are evaluated by \verb|MadGraph2.5.3_aMC@NLO|.  We calculated  the FCNC decay widths $\Gamma(t\to q\gamma)$ and $\Gamma(t\to q g)$ depending on coupling $\lambda_q$ and $\zeta_q$ is defined as 
\begin{eqnarray}\label{width}
\Gamma(t\to q\gamma)&=&0.6723 \lambda_q^2\\
\Gamma(t\to q g)&=&49.55 \zeta_q^2
\label{decay}
\end{eqnarray}
Using Eqs. (4) and (5) and total decay width of the top quark $\Gamma(t \to Wb)=1.47$, the FCNC coupling $\lambda_q$=0.0027 and $\zeta_q=0.0018$ obtained from  Fig.~\ref{ss_two} at $2\sigma$ SS value can be converted to the branching ratio $BR(t\to q\gamma)=3.3\times10^{-6}$  and $BR(t\to qg)=1.1\times10^{-4}$ for hadronic channel with L$_{int}$=10 ab$^{-1}$. These branching ratios are at the same order for leptonic channel. 

By comparing different formulations of the anomalous FCNC top couplings, the effective dimension-5 $tq\gamma$ and $tqg$ operators and SM gauge invariant dimension-6 operators (with couplings $c_i$) \cite{AguilarSaavedra:2009mx}, the new physics cut-off scale $\Lambda$ can be expressed as $\Lambda (\lambda_q) / \sqrt {c_i}= \sqrt{\sqrt 2 v m_t(\sin{\theta_W} + \cos{\theta_W})/g_e\lambda_q}$ and $\Lambda(\zeta_q)/\sqrt {c_i}  =\sqrt{\sqrt 2 v m_t/g_s\zeta_q}$, where $v$ is vacuum expectation value, $\theta_W$ is the Weinberg angle. Assuming $c_i =\mathcal O (1)$, we calculate $\Lambda$=9.8 TeV and 5.2 TeV  for the limits on $\lambda_q$=0.0027 and $\zeta_q=0.0018$, respectively. 

We compare our results on the branching ratios with the current experimental results summarized in Table \ref{BR}. Based on proton-proton collisions at 8 TeV within the CMS detector at the LHC at an integrated luminosity of 19.8 fb$^{-1}$, the limits on the top quark FCNC branching ratios are $BR(t\to u\gamma)=1.7\times10^{-4}$ and $BR(t\to c\gamma)=2.2\times10^{-3}$ at 95\% C.L. \cite{Khachatryan:2015att}. Our limit on the branching ratio for $t \to q \gamma$  is  three order smaller than the current CMS experimental results and one order better than the projected limits on top FCNC couplings through top pair production in which one of the top quark decays to $Wb$ while the other decays to $q\gamma$ at LHC 14 TeV and HL-LHC reported in Ref. \cite{ATLAS:2013hta}, where the expected upper limits on branching ratio $t\to q\gamma$ are $8\times10^{-5}$ and $2.5\times10^{-5}$ for an integrated luminosity 300 fb$^{-1}$ and 3000 fb$^{-1}$, respectively. The CMS projections for 14 TeV pp collisions with an integrated luminosity of 3000 fb$^{-1}$ using CMS Delphes simulation for the sensitivity of FCNC $t q\gamma$ coupling through the processes $pp\to t\gamma$ and $pp\to t\bar t$ with one of the top quark decaying via FCNC is presented for branchings $BR(t\to u\gamma)=2.7\times10^{-5}$ and $BR(t\to c\gamma)=2.0\times10^{-4}$ at 95\% C.L \cite{CMS:2017gvo}. We also compare our results for the projections of FCNC $t q\gamma$ couplings at 14 TeV and 100 TeV. We use the calculated signal cross sections $\sigma$=0.8528 pb  at 14 TeV LHC and $\sigma$=8.528 pb at 100 TeV FCC-hh for the process $pp\to t\gamma$ with the equal couplings scenario ($\lambda_q$=0.01 and $\zeta_q$=0.01). An order of magnitude enhancement in signal cross section is due to the higher center of mass energy. For the SM background with the same final state as signal process $pp\to W b \gamma$ the cross sections  $\sigma$=0.02506 pb  at 14 TeV LHC and  $\sigma$=0.3831 pb at 100 TeV FCC-hh. A comparison of these enhancements in the cross section and in the luminosity (3 ab$^{-1}$ for HL-LHC and 10 ab$^{-1}$ for FCC-hh) can also be converted into the limits on branching ratios therefore an order order of magnitude better sensitivity can be explained. 

The limits on the FCNC branching ratio of the $t\to q g$ decay channel are $BR(t\to u g)=4.0\times10^{-5}$ and $BR(t\to cg)=20\times10^{-5}$ reported from ATLAS collaboration via single top-quark production with flavor-changing neutral current processes in proton-proton collisions at a centre-of-mass energy of 8 TeV and corresponding to an integrated luminosity of 20.3 fb$^{-1}$. The limit on the $BR(t\to qg)$ with the anomalous single top quark production in association with a photon process ranges at the same order as current ATLAS experiment. 
\section{Conclusions}
Deviations from the SM predictions are often interpreted in terms of anomalous top couplings in the single top production. One can put constraints on each effective operators which could describe these possible deviations. In this paper, the anomalous top FCNC couplings of  $tq\gamma$ and $tqg$ vertices in a model independent way have been investigated via $pp\to Wb\gamma$ signal process at 100 TeV center of mass energy. The both leptonic and hadronic decay channels of W boson in the final state of the signal are taken into account to obtain sensitivities of the anomalous couplings at FCC-hh including realistic detector effects in the analysis. Using distinctive feature of the prompt photon radiation in the final state of the signal process, the top FCNC interactions can be uncovered from overwhelming relevant SM backgrounds. Thus, high $p_T^{\gamma}$ cut with other optimum kinematic cuts requirement are used  as a tool to probe sensitivity of the anomalous couplings. With an integrated luminosity of 10 ab$^{-1}$ and $\sqrt s$= 100 TeV for a 2$\sigma$ SS value, the sensitivity to the branching ratio of $t \to q \gamma$ channel is three order better than the available experimental limits, and comparable  for the branching ratio of the $t\to qg$ decay channel.

\begin{acknowledgments}
This work was supported by Turkish Atomic Energy Authority (TAEK) under the grant No. 2018TAEK(CERN)A5.H6.F2-20. We  acknowledge exciting discussion within the FCC-hh physics analysis meeting. The K.Y. O., A. S. and H. D. work partially supported by the Bolu Abant Izzet Baysal University Scientific Research Projects under the Project no: 2018.03.02.1286.
\end{acknowledgments}

\newpage
 \begin{table}
\caption{The current experimental 95\% C.L. upper limits on the branching fractions of the top quark FCNC decays obtained at the LHC experiments.  \label{BR}}
\begin{ruledtabular}
\begin{tabular}{cccc}
Decay Channels&$BR(q=u)$&$BR(q=c)$& Ref. \\ \hline 
$t\to q g$ & 4.0$\cdot 10 ^{-5}$ &2.0$\cdot 10 ^{-4}$&\cite{Aad:2015gea} \\
$t\to q \gamma$ &   1.3$\cdot 10 ^{-4}$&1.7$\cdot 10 ^{-3}$&\cite{Khachatryan:2015att}\\
$t \to q Z$&  2.2$\cdot 10 ^{-4}$ &4.9$\cdot 10 ^{-4}$& \cite{Sirunyan:2017kkr}\\
$t \to q H$ & 2.4$\cdot 10 ^{-3}$&2.2$\cdot 10 ^{-3}$&\cite{Aaboud:2017mfd}\\
\end{tabular}
\end{ruledtabular}
\end{table}

\begin{table}
\caption{Event selection and kinematic cuts used for the analysis of signal and background events in hadronic and leptonic channels. \label{tab1}}
\begin{ruledtabular}
\begin{tabular}{lccc}
Cuts & Leptonic channel &Hadronic channel \\ \hline 

Pre-selection & $N_\gamma \geqslant1$, $N_l \geqslant 1$ and $N_b = 1$&$N_\gamma \geqslant1$, $N_j \geqslant 3$ , $N_b = 1$ and no lepton  \\
Kinematic-I (II)& $p_{T}^{\gamma} > 50 (150)$ GeV, &$p_{T}^{\gamma} > 50 (150)$ GeV\\
&  $p_{T}^{l} > 30$ GeV, $p_{T}^b > 30$ GeV&$p_{T}^{j1,j2} > 30$ GeV, $p_{T}^b > 30$ GeV\\
& $|\eta^{\gamma,l,b}|<2.5$, MET > 30 GeV  & $|\eta^{\gamma,j1,j2,b}| < 2.5$, $\Delta R (b,\gamma)$ > 0.7 \\
& $\Delta R (b,\gamma)$ > 0.7 and $\Delta R (l,\gamma)$ > 0.7&$\Delta R (j1,\gamma)$ > 0.7 and $\Delta R (j2,\gamma)$ > 0.7 \\
W-reconstruction & 80 GeV < $m_{l\nu}$ < 90 GeV & 35 GeV < $m_{jj}$ < 90 GeV\\
Top-reconstruction & 135 GeV < $m_{l\nu b}$ < 195 GeV & 135 GeV < $m_{jjb}$ < 195 GeV \\

\end{tabular}
\end{ruledtabular}
\end{table}

\begin{table}
\caption{The number of signal and relevant background events after each kinematic cuts in the analysis single lepton mode with $L_{int}=$100 fb$^{-1}$.\label{tab2} }
\begin{ruledtabular}
\begin{tabular}{lcccccc}
Processes& Pre-selection&Kinematic-I&Kinematic-II& W-reconstruction&Top-reconstruction& \\
\hline 
Signal ($\lambda_q=0.01$)& 11338 &3229& 2478 & 2175 & 1365    \\
Signal($\zeta_q=0.01$)& 303576 &15138& 6620 & 6039&  3534  \\
Signal($\lambda_q=0.01$,$\zeta_q=0.01$)&319686 &21628& 12283  & 10602 & 7195   \\
sm &2584 &107 &18 &14 & 3   \\
$Wj\gamma$&657203 &24091& 5815  & 5192 & 1038 \\
$Wj$& 1.679$\cdot10^8$&43630 &0& 0 & 0  \\
$tt$& 3.476$\cdot10^8$&800428& 83589 & 37995 & 27863   \\
$tt\gamma$& 1.636$\cdot10^6$& 215933& 70488 &  50583 &   22560   \\
$Zj\gamma$& 465056 &2054& 313 &  244 &   70   \\
\end{tabular}
\end{ruledtabular}
\end{table}

\begin{table}
\caption{The number of signal and relevant background events after each kinematic cuts in the analysis full hadronic mode with $L_{int}=$100 fb$^{-1}$.\label{tab3} }
\begin{ruledtabular}
\begin{tabular}{lccccccc}
Processes& Pre-selection&Kinematic-I&Kinematic-II& W-reconstruction&Top-reconstruction& \\
\hline 
Signal ($\lambda_q=0.01$)& 15214 &7986& 5616 & 2349 & 1740   \\
Signal($\zeta_q=0.01$)& 381507 &43778& 15059  &5776&  4035   \\
Signal($\lambda_q=0.03$,$\zeta_q=0.01$)&411290 &61438& 28782  & 11836 & 8771  \\
sm &3226.04 &311 &48 &21 & 5   \\
$Wj\gamma$&1.23$\cdot10^6$ & 130319&23260  & 4673 & 1246  \\
$Wj$& 3.182$\cdot10^8$&261780 & 43630& 0 & 0  \\
$tt$& 3.502$\cdot10^8$& 2.01$\cdot10^6$&169711 &   58259 & 20264   \\
$tt\gamma$& 1.315$\cdot10^6$ & 272273&81031 & 16464 &   7014   \\
$Zj\gamma$& 1.318$\cdot10^6$ & 128103&19987 &  3935 &   1637  \\
$jjjj$&5.567$\cdot10^{11}$ & 1.33$\cdot10^{8}$& 4.459$\cdot10^7$&   0 & 0  \\
$bjjj$& 9.953$\cdot10^{10}$ & 2.02$\cdot10^{7}$&  0 &   0 &  0   \\
\end{tabular}
\end{ruledtabular}
\end{table}
\newpage

\end{document}